\documentclass[11pt,a4paper]{article}
\usepackage{graphicx}
\usepackage{titlesec}
\usepackage{titling}
\usepackage{tabularx}
\usepackage[utf8]{inputenc}
\usepackage[T1]{fontenc}
\usepackage[hmargin=2.5cm,vmargin=3cm]{geometry}
\usepackage{amsmath,amssymb,amsfonts}
\usepackage{mathtools}
\usepackage{libertinus}
\usepackage{libertinust1math}
\usepackage[svgnames]{xcolor}
\usepackage{hyperref}
\hypersetup{colorlinks=true, linkcolor=NavyBlue, citecolor=DarkRed, urlcolor=blue}
\usepackage{lineno}
\usepackage{booktabs}
\usepackage{array}
\usepackage[font=small,labelfont=bf]{caption}
\usepackage{tikz}
\usepackage[mathscr]{eucal}
\usepackage{dutchcal} 
\usepackage{bm}
\allowdisplaybreaks
\usepackage{comment}
\usepackage{makecell}
\usepackage{accents}
\usepackage{appendix}

\usepackage{relsize} 

\usepackage[bb=boondox]{mathalfa} 
\usepackage[backend=bibtex,giveninits=true,sorting=nyt,url=false,doi=true,eprint=false,isbn=false,
backref,backrefstyle=none,maxbibnames=99]{biblatex}
\DefineBibliographyStrings{english}{%
	backrefpage = {Cited on p\adddot},%
	backrefpages = {Cited on pp\adddot}%
}
\bibliography{references}

\renewbibmacro{in:}{%
	\ifboolexpr{%
		test {\ifentrytype{article}}%
	}{}{\printtext{\bibstring{in}\intitlepunct}}%
}

\newcommand\BibTeX{{\rmfamily B\kern-.05em \textsc{i\kern-.025em b}\kern-.08em
T\kern-.1667em\lower.7ex\hbox{E}\kern-.125emX}}

\allowdisplaybreaks[1]

\usepackage{hyperref}
\hypersetup{
    colorlinks=true,
    linkcolor=Blue,
    citecolor=DarkRed,
    urlcolor=blue  }

\newcommand{\pd}{\partial}

\newcommand{\rmd}{\mathrm{d}}
\newcommand{\XX}{\bm{X}}

\newcommand{\gL}[2]{G_{#1#2}}          
\newcommand{\gE}[2]{g_{#1#2}}          
\newcommand{\Tors}[3]{\mathbb{T}^{#1}_{\ #2#3}}  

\newcommand{\F}[1]{\mathbb{F}_{#1}}  
\newcommand{\FStar}[1]{{}^\star\hspace{-0.3mm}\mathbb{F}^{#1}}
\newcommand{\EF}[1]{\mathbb{E}_{#1}}
\newcommand{\DF}[1]{\mathbb{D}^{#1}}
\newcommand{\BF}[1]{\mathbb{B}^{#1}}

\newcommand{\Hexci}[1]{\mathbb{H}^{#1}}
\newcommand{\Ch}{\varkappa_1}	
\newcommand{\ch}{\varkappa_2}	

\newcommand{\EFvec}{{\boldsymbol{\mathbb{E}}}}
\newcommand{\DFvec}{{\boldsymbol{\mathbb{D}}}}
\newcommand{\BFvec}{{\boldsymbol{\mathbb{B}}}}
\newcommand{\HFvec}{{\boldsymbol{\mathbb{H}}}}
\newcommand{\JDmicrovec}{\boldsymbol{\psi}}
\newcommand{\JDtotvec}{\boldsymbol{\Psi}}
\newcommand{\qvec}{{\boldsymbol{q}}}
\newcommand{\Tprop}{\hat{\Theta}}
\newcommand{\Ttot}{\Theta}
\newcommand{\Tmicro}{\theta}
\newcommand{\Tlarge}{\Xi}
\newcommand{\Stot}{S}

\newcommand{\Smicro}{s}
\newcommand{\Jlargevec}{\bm{\mathscr{J}}}
\newcommand{\Jlarge}{\mathscr{J}}
\newcommand{\Jmicrovec}{\bm{A}}
\newcommand{\Jmicro}{A}
\newcommand{\Jtotvec}{{\bm{J}}}
\newcommand{\Jtot}{J}
\newcommand{\Asm}{{\mathsmaller{A}}}
\newcommand{\Bsm}{{\mathsmaller{B}}}
\newcommand{\Csm}{{\mathsmaller{C}}}
\newcommand{\Dsm}{{\mathsmaller{D}}}

\newcommand{\Osm}{{\mathsmaller{O}}}
\newcommand{\Ksm}{{\mathsmaller{K}}}
\newcommand{\Lsm}{{\mathsmaller{L}}}
\newcommand{\Ism}{{\mathsmaller{I}}}
\newcommand{\Jsm}{{\mathsmaller{J}}}
\newcommand{\Tsm}{{\mathsmaller{T}}}
\newcommand{\Jtotcurl}{{\bm{\Omega}}}

\newcommand{\Vph}{V_\text{ph}}
\newcommand{\Vgr}{V_\text{gr}}

\newcommand{\Ham}{\mathcal{H}}

\newcommand{\veps}{\varepsilon}
\newcommand{\eps}{\epsilon}
\newcommand{\trans}{\mathsf{T}}
\newtheorem{remark}{Remark}

\begin{document}

\title{ \Large
	\textbf{Geometrization of continuum mechanics and nonequilibrium thermodynamics: Heat conduction}
}
\author{
Ilya Peshkov\thanks{Department of Civil, Environmental and Mechanical Engineering,
University of Trento, Via Mesiano 77, Trento 38123, Italy
\href{mailto:ilya.peshkov@unitn.it}{ilya.peshkov@unitn.it}
}
}
\maketitle 
\date{\today}

\begin{flushright}
	\small
\emph{In memory of Miroslav Grmela.}
\end{flushright}

\begin{abstract}
\noindent
We propose a geometric formulation of non-equilibrium heat conduction. It
combines the variational (Hamiltonian) structure of symmetric hyperbolic
thermodynamically compatible (SHTC) equations with the exterior calculus of
differential forms. The central object is the thermal impulse one-form on the 4D
matter-time manifold. Its time component is the temperature, and for reversible
(ballistic) heat transport it is the exact differential of a scalar potential
(i.e. it is holonomic). Non-equilibrium effects are described by a second
one-form, the small-scale thermal impulse, which is not required to be exact.
Its time component is a fluctuating temperature, and its exterior derivative,
the four-thermal vorticity, satisfies a Bianchi-type kinematic identity. The
variational principle together with this identity yields a non-equilibrium
two-temperature model. Its field equations have the same structure as massive
Maxwell-Proca-type gauge theories, with the thermal impulse field, the thermal
vorticity field and the thermal Poynting vector contributing to the heat flux.
Dissipation is introduced through relaxation terms that conserve the total
energy and produce entropy. A formal asymptotic analysis shows that, in the
relaxation limit, the model reduces to a Guyer-Krumhansl-type heat conduction
with independent transverse and longitudinal length scales. The transverse
nonlocal term comes from the spatial non-holonomy of the small-scale thermal
impulse, while the longitudinal term comes from its non-holonomy in time, i.e.
from the relaxation of the fluctuating temperature. We also discuss the
gauge-theoretic interpretation of the model. Its kinematics is that of an
abelian gauge theory, while the local gauge symmetry is explicitly broken by the
existence of two temperatures. Finally, we discuss the consequences of the
geometric structure for structure-preserving numerical methods.

\end{abstract}


\section{Introduction}

It is an old and still somewhat unexplained observation that the most successful
physical theories are, almost without exception, theories with a geometric
structure, e.g. Riemannian geometry behind the general relativity, fiber bundle
geometry behind gauge theories (electrodynamics, standard model of particle
physics). Typically, such a geometric structure is not apparent in the equations
as first written down; it is usually discovered afterwards (except maybe the
general relativity).  However, once discovered, it allows the theory to be
generalized, coupled to other theories, and discretized preserving its important
structural properties \cite{ArnoldFalkWinther2006}.

In this paper, we propose a geometric formulation of the heat conduction. This,
however, results in the need to abandon two important features of the classical
Fourier theory. Namely, we will not assume the local-equilibrium hypothesis and
we will work in the class of hyperbolic partial differential equations (PDEs)
rather than parabolic ones. The former is required because, as any equilibrium
theory, the Fourier model is a reduced description (i.e. in a reduced phase
space) and does not allow one to see geometric structures hidden in the higher
dimensional phase spaces. Furthermore, it has appeared that our geometric
formulation has to be a causal theory, and hence to be built in the hyperbolic
class of PDEs, because in the proposed framework, this is the reversible (and
hence causal) part of the time evolution that carries the differential-geometric
structure. Yet, we stress that the diffusive Fourier heat conduction is
recovered as a particular relaxation limit (in some small parameter) of the
proposed theory.

Two families of geometric structures are worth recalling, because the proposed
theory of heat conduction borrows from both of them. The first is the
\emph{symplectic and Poisson} structure of Hamiltonian mechanics
\cite{Arnold1989,MarsdenRatiu1999} which realizes as the contact geometry of jet
bundles, and in the context of non-equilibrium thermodynamics is realized as its
dissipative extensions in the GENERIC framework by Grmela, \"Ottinger and
collaborators
\cite{GrmelaOttingerI,GrmelaOttingerII,Ottinger2005,PKG_Book2018,SHTC_GENERIC},
or metriplectic formalism by Morrison
\cite{Morrison1984,Morrison1986,Morrison2024}. The second is the
differential-geometric structure of \emph{exterior calculus} of differential
forms, which underlies the coordinate-free formulation of electrodynamics, of
the theory of continuously distributed defects, and of general relativity
\cite{Frankel2011,AldrovandiPereiraBook,MTW1973,Kroner1981,Clayton2005,Steinmann2015,LychevKoifman2018},
and which has important feature that the structural identities of a theory
appear as instances of the vanishing second exterior derivative
\begin{equation*}
	\rmd^2 \omega = 0
\end{equation*}
for any differential $k$-form $\omega$.

To see how these two geometric frameworks are combined in the proposed theory,
let us remark that we shall work in a particular class of hyperbolic equations,
the so called symmetric hyperbolic thermodynamically compatible (SHTC) class of
PDEs developed by Godunov and Romenski in
\cite{God1961,GodRom1996,GodRom1998,SHTC_GENERIC}. An important feature of the
SHTC class for the proposed geometric formulation is that the governing
equations are organized in pairs of a \emph{dynamical} equation (e.g. the
electric field PDE in electrodynamics) and a \emph{kinematic} identity (e.g. the
magnetic field PDE in electrodynamics). The first is obtained from a variational
principle, and hence carries the symplectic structure of the theory, while the
second is a purely kinematic identity that follows from the
differential-geometric structure of the fields and holds independently of the
dynamics, i.e. as an instance of $\rmd^2 \omega = 0$.

The central object of the presented geometric formulation of heat conduction is
a $1$-form, \emph{thermal impulse}, on the matter--time manifold with
coordinates $X^\Asm$, introduced in the previous realization of the presented
theory \cite{RomenskiMalyshev1987,PTRSA2020,SHTC_GENERIC},
\begin{equation*}
  \Jtotvec = \Jtot_{\Asm} \rmd X^\Asm ,
  \qquad
  \hat{\Theta}=\sqrt{ -\Jtot_\Asm J^\Asm },
\end{equation*}
whose Lorentzian norm is the proper temperature $\hat{\Theta}$ and whose components
generalize the thermal displacement gradient of Green--Naghdi type theories
\cite{GreenNaghdi1991,GreenNaghdi1993,vanDantzig1939}. In the reversible limit
(e.g. ballistic transport), $\Jtotvec$ can be approximated as a closed form and can
be written as $\Jtotvec = \rmd\varphi$; but once the dissipation is present,
$\Jtotvec$ is not closed, and the $2$-form
\begin{equation*}
  \F{} = \rmd\Jtotvec,
  \qquad
  \F{\Asm\Bsm} = \pd_\Asm \Jtot_\Bsm - \pd_\Bsm \Jtot_\Asm,
\end{equation*}
which we call here the \emph{thermal vorticity} following \cite{Sykora2023},
becomes the carrier of the non-equilibrium content of the theory. Everything
that follows from these initial postulates (entropy conservation law,
Maxwell-like structure of equations, etc) is a consequence of this single
kinematic choice together with the identity $\rmd^2 \omega = 0$.

As it was mentioned above, our geometric formulation is intrinsically a
non-equilibrium theory of heat conduction, it is therefore important to properly
place it within the context of existing non-equilibrium models. A considerable
number of non-equilibrium thermodynamic theories have been proposed over the
years, several of them specifically for heat conduction. As is customary, they
can be organized by the order at which they truncate an expansion in a small
parameter measuring the departure from local equilibrium (e.g. the Knudsen
number, or, equivalently, the number of moments retained beyond the Fourier
closure). At the 0-th order (\emph{diffusive regime}), the heat conduction is
well described by the classical Fourier model. At the 1-st order (\emph{second
sound}), we include the Cattaneo-Vernotte-type theories
\cite{Cattaneo1948,Vernotte1958}. The most developed are Rational Extended
Thermodynamics (RET) of M\"uller and Ruggeri \cite{MullerRuggeri1998}, Extended
Irreversible Thermodynamics (EIT) of Jou, Casas-V\'azquez and Lebon
\cite{JouCasasLebon2010}, moments-based theories, e.g. by Dreyer and Struchtrup
\cite{dreyer1993}, the relativistic moments theories of Israel and Stewart
\cite{IsraelStewart1979} and of Carter \cite{Carter1989}, or by Anderson and
Comer \cite{AnderssonComer2015}, and the SHTC and GENERIC formulations
\cite{RomenskiMalyshev1987,SHTC_GENERIC,Dhaouadi2024}, and
\cite{Ottinger1998,Ottinger1998b,Sykora2023,Ottinger2005,PKG_Book2018}. All of
these promote the heat flux, or the thermal impulse, to an independent field
governed by its own balance law, and in this sense are hyperbolic
generalizations of Cattaneo-Vernotte relaxation model
\cite{Cattaneo1948,Vernotte1958}. The Green--Naghdi type-II/III theories
\cite{GreenNaghdi1991,GreenNaghdi1993,vanDantzig1939} belong to the same
first-order family, formulated instead in terms of the thermal displacement. At
the 2-nd order, the theories must include the diffusive phonons, second sound,
and additionally \emph{ballistic phonons}. The reference models at this order
are the Guyer--Krumhansl (GK) equation \cite{Guyer1966}, derived from kinetic
theory and applied to nanoscale and room-temperature heat transport
\cite{Van2017,Zhu2017,JouCasasLebon2010}, and the RET-type theories
\cite{MullerRuggeri1998}, see for example the theory by Dreyer and Struchtrup
\cite{dreyer1993} and references therein. More recently, motivated by phonon
hydrodynamics and thermal vortices \cite{Sykora2023,dilucente2026},
generalizations of the GK equation have been obtained by adding internal
variables that describe the viscous and the vortical (rotational) motion of the
phonon flow, both within EIT with internal variables and within GENERIC
\cite{restuccia2023,restuccia2026}. Note that RET-type theories can be
extended to higher-order models, due to their hierarchical structure as more and
more moments of the phonon distribution function can be potentially retained in
a macroscopic model.

A different and complementary route is followed by phenomenological theories
based on the Clau\-sius-Duhem inequality and the Coleman--Noll procedure. A
recent example is the framework of Giorgi and Zullo \cite{giorgi2026}, see also
the references therein. There, the free energy and the entropy production depend
on the temperature, its gradient, the heat flux and their time derivatives, and
the second law of the thermodynamics is imposed in a (weakly) nonlocal form with
an entropy extra-flux. Within such theoreis, it is possible to recovers a large
family of rate-type heat conductors: Maxwell-Cattaneo, Jeffreys-like,
Green-Naghdi, Burgers-like and, in the weakly nonlocal case, Guyer-Krumhansl and
its nonlinear generalizations. Theories of this type are in a sense more general
than the one proposed here, because the admissible constitutive relations are
restricted only by the entropy inequality. 

None of the above mentioned theories, however, comes with a
differential-geometric structure: the extra fields they introduce are treated as
ordinary vectors or tensors on space(time), not as sections of a bundle subject
to curvature or torsion constraints. The absence of a geometric structure is not
a drawback per se: there is no fundamental evidence that non-equilibrium
thermodynamics must possess one, and the thermodynamic-consistency approach has
proven very successful. Nevertheless, we believe that such a structure brings
apparent benefits, both for the physical interpretation and for the numerical
treatment of the models. We summarize these arguments in
Section~\ref{sec:discussion}. The model presented in this paper can be viewed as
such a geometric completion of the non-equilibrium heat conduction. As we shall
see, the minimal setting required to introduce the necessary geometric structure
(e.g. the curvature tensor) puts us already at the level of 2-nd order models in
the above classification (or weakly non-local models in terms of
\cite{giorgi2026}). In particular, we show in Section~\ref{sec:GK2} that
the GK-type equation is recovered as a particular relaxation limit of the theory
developed here.

For simplicity, in this paper, we assume that the matter--time manifold
$\mathscr{M}$ is a flat 4D Lorentzian manifold. However, the power of the
geometric methods, and in particular of exterior calculus, is that they can be
applied to manifolds of arbitrary geometry, and in particular to curved
manifolds, see Section\,\ref{rem:torsion} below.

The paper is organized as follows. Section~\ref{sec:conventions} fixes the
notation. In Section~\ref{sec:base}, we recall the base (one-temperature) model
of heat conduction, its variational formulation, and its dissipative extension.
Section~\ref{sec:ext.model} introduces the two-temperature model. We first give
its geometric set-up and variational formulation, then its $3+1$ form with
Maxwell-type structure, and finally its dissipative extension and the hierarchy
of models it contains. In Section~\ref{sec:GK2} we show by a formal asymptotic
expansion that the GK-type equation is a relaxation limit of the two-temperature
model. The dispersion relation of the two-temperature model is investigated in
Section~\ref{sec:dispersion}. Finally, Section~\ref{sec:discussion} discusses
the physical context of the new fields, the gauge-theoretic interpretation of
the model, its bundle-geometric generalizations, and the implications for
numerical discretization.

\section{Definitions and conventions}\label{sec:conventions}

We assume that the space-time manifold $\mathscr{E}$ is a 4D pseudo-Euclidean
manifold with a flat Minkowski metric $\gE{\mu\nu}{}=\text{diag}(-1,1,1,1)$ in a
coordinate system $x^\mu$, $\mu=0,1,2,3$. We treat the time evolution of the
material manifold as an \textit{embedding} of a 4D matter-time manifold
$\mathscr{M}$ into the space-time manifold $\mathscr{E}$. The intrinsic geometry
of $\mathscr{M}$ is potentially arbitrary. However, in this paper we restrict
ourselves to the case of a flat material manifold with a metric
$\gL{\Asm\Bsm}{}=\text{diag}(-1,1,1,1)$ in the material coordinates $X^\Asm $,
$\Asm,\Bsm,\Csm,\ldots=\Osm,1,2,3$, with $X^\Osm$ being the time coordinate, and
$X^\Ksm$, $\Ksm=1,2,3$ being the spatial coordinates, with
$\Ism,\Jsm,\Ksm=1,2,3$ reserved for the spatial indices.

We also will need the 4D and 3D Levi-Civita symbols $\veps^{\Asm\Bsm\Csm\Dsm}$
and $\veps^{\Ism\Jsm\Ksm}$, for which we adopt the following sign convention:
\begin{equation}
	\veps_{0123} =-1,
	\qquad
	\veps^{\Osm\Ism\Jsm\Ksm} = \veps^{\Ism\Jsm\Ksm},
	\qquad
	\veps_{\Osm\Ism\Jsm\Ksm} = -\veps_{\Ism\Jsm\Ksm}.
\end{equation}

\section{Base model}\label{sec:base}

We now discuss the base model of heat conduction proposed by Romenski and
Malyshev in \cite{RomenskiMalyshev1987} (see also
\cite{PTRSA2020,SHTC_GENERIC}), which does not enjoy a complete geometric
structure but it worth recalling it as it serves as a starting point for the
geometric formulation which will be developed in the subsequent section. But
before proceeding, we need to briefly discuss two inconsistencies of the
classical phenomenological Fourier theory, which in turn were the motivation for
the base model and subsequent geometric formulation. As we already mentioned in
the introduction there are two interrelated issues that we need to overcome.
First, the classical Fourier theory lacks a variational formulation, which would
give us a multisymplectic structure of contact geometry. Second, it lacks a
reversible kinematic structure (e.g. like the magnetic field PDE in
electrodynamics), which would give us a differential-geometric structure of
exterior calculus.

Indeed, the classical Fourier theory (as any phenomenological equilibrium
transport theory, e.g. Fickian diffusion, Darcy's law of filtration, Newton's
law of viscosity, etc.) relies on the assumption that the local equilibrium can
be established \textit{instantaneously}, which introduces the known causality
issue: the paradox of infinite speed of propagation of heat, e.g. see
\cite{MullerRuggeri1998,JouCasasLebon2010}, as well as the discussions in
\cite{Bright2009,Sobolev2022}. Such a removal of inertial effects at the
microscopic level, results in a completely dissipative/irreversible dynamics at
the macroscopic level, which mathematically is expressed in the parabolic energy
conservation law. Therefore, if one wants to have a variational formulation of
the theory and a differential-geometric structure, one has to introduce
reversibility into the time evolution of the heat propagation process. This
means that we split the time evolution into a reversible and an irreversible
part, e.g. in the spirit of the GENERIC
\cite{GrmelaOttingerI,GrmelaOttingerII,PKG_Book2018,SHTC_GENERIC}, or
metriplectic \cite{Morrison1984,Morrison2024} formulations of the nonequilibrium
thermodynamics. However, such a split is not simply about modifying the elliptic
constitutive relation of the Fourier theory, but it requires introducing new
degrees of freedom into the theory (e.g. independent heat flux), which would
represent the microinertia missing in the classical equilibrium theories.

The base model \cite{RomenskiMalyshev1987,PTRSA2020,SHTC_GENERIC} already solved
the first issue by introducing a variational formulation of the theory, and
hence enjoying the multisymplectic structure of contact geometry, while the
extension, which allows to couple the two geometric frameworks (contact geometry
and exterior calculus) is the main subject of this paper, and is presented in
Section\,\ref{sec:ext.model}.

\subsection{Variational formulation of the base model}
We consider the action integral of the form:
\begin{equation}
	S[\varphi] = \int \Lambda(\varphi,\varphi_{,\Asm}) \rmd^4 \XX
\end{equation}
where $\varphi(X^\Asm )$ is a scalar potential (sometimes called the thermal
displacement after Green-Naghdi \cite{GreenNaghdi1991}), $\varphi_{,\Asm} :=
\frac{\partial \varphi}{\partial X^\Asm }$, and $\Lambda$ is the Lagrangian
density.

The Euler-Lagrange equation that follows from the variational principle, i.e.
from the stationarity of the action integral $\delta S/\delta \varphi = 0$, is
\begin{equation}\label{eq:EL1}
	\frac{\delta S}{\delta \varphi} =
	\frac{\partial \Lambda}{\partial \varphi}
	- \frac{\partial}{\partial X^\Asm } 
	\left( \frac{\partial \Lambda}{\partial \varphi_{,\Asm}} \right) = 0.
\end{equation}

In what follows, we will assume that the Lagrangian density $\Lambda$ does not
depend explicitly on the potential $\varphi$, i.e. the Euler-Lagrange equations
reduce to the conservation law:
\begin{equation}\label{eq:EL2}
	\pd_\Asm \Lambda_{\varphi_{,\Asm}} = 0,
\end{equation}
where we use the shorthand notations $\pd_\Asm = \frac{\partial}{\partial X^\Asm
}$ and $\Lambda_{\varphi_{,\Asm}} = \frac{\partial \Lambda}{\partial
\varphi_{,\Asm}}$.

Equation \eqref{eq:EL2} is a second-order nonlinear wave equation on the
potential $\varphi$ which is a hyperbolic PDE. However, for the geometric
formulation developed later, we will consider it together with the identity
(\emph{integrability condition}) $\pd_\Asm \varphi_{,\Bsm} - \pd_\Bsm
\varphi_{,\Asm} = 0$ as a system of first-order PDEs on the four-vector field
$\varphi_{,\Asm}$:
\begin{equation}\label{eq:EL3}
	\pd_\Asm \Lambda_{\varphi_{,\Asm}} = 0, \qquad
	\pd_\Asm \varphi_{,\Bsm} - \pd_\Bsm \varphi_{,\Asm} = 0.
\end{equation}

In 3D notations, the system \eqref{eq:EL3} is equivalent to the following time
evolution equations:
\begin{equation}\label{eq:EL4}
	\pd_\Tsm \Lambda_{\varphi_{,\Osm}} 
	+ \pd_\Ksm  \Lambda_{\varphi_{,\Ksm}} = 0,
	\qquad
	\pd_\Tsm \varphi_{,\Ksm} - \pd_\Ksm  \varphi_{,\Osm} = 0,
\end{equation}
plus three more stationary constraints $\pd_\Lsm \varphi_{,\Ksm} - \pd_\Ksm
\varphi_{,\Lsm} = 0$. Note that $\varphi_{,\Osm} = \pd_\Tsm \varphi$ is the time
derivative of the potential $\varphi$.

For the following discussion, we introduce the notation
\begin{equation}\label{eq:JA}
	\Jtot_\Asm := \pd_\Asm\varphi = \varphi_{,\Asm},
\end{equation}
with a possibility for $\Jtot_\Asm$ to depart from its definition \eqref{eq:JA}
as a gradient of the potential $\varphi$ in the presence of dissipation later.
We shall call $\Jtot_\Asm$ the \textit{thermal impulse} and interpret it as the
momentum of the heat carriers (phonons), which is similar to other works on
non-equilibrium heat conduction, e.g. see the work by Dreyer and Struchtrup
\cite{dreyer1993}.

We then single out the time component of $\Jtot_\Asm = (-\Ttot, \Jtot_\Ksm )$:
\begin{equation}
	\Ttot:=-\varphi_{,\Osm} =-\pd_\Tsm \varphi,
	\qquad
	\Jtot_\Ksm  := \varphi_{,\Ksm} = \pd_\Ksm  \varphi, \quad K=1,2,3,
\end{equation}
where $\Ttot$ is interpreted as the temperature\footnote{One may also to
consider the \textit{proper temperature}: $\Tprop = \sqrt{-\Jtot_\Asm J^\Asm } =
\sqrt{\Ttot^2 - \Jtot_\Ksm J^K}, $ so that in the frame co-moving with the heat
carriers, i.e. $\Jtot_\Ksm  = 0$, we have $\Tprop = \Ttot$.}, and consider the
Hamiltonian density $\Ham(\Stot,\Jtot_\Ksm )$ as the partial Legendre transform
of the Lagrangian density $\Lambda(\Ttot,\Jtot_\Ksm )$ with respect to the time
component of the thermal impulse $\Jtot_\Asm$:
\begin{equation}
	\Ham:= \Ttot \Lambda_{\Ttot} - \Lambda
\end{equation}
The entropy is then defined as the dual variable to the temperature:
\begin{equation}
	\Stot := \Lambda_{\Ttot},
	\qquad
	\Ttot = \Ham_\Stot,
\end{equation}
After that, system \eqref{eq:EL4} becomes a system on the phase space $(\Stot,
\Jtot_\Ksm )$ and can be rewritten in the canonical SHTC form
\cite{RomenskiMalyshev1987,GodRom1996,SHTC_GENERIC}:
\begin{equation}\label{eq:EL5}
	\pd_\Tsm \Stot + \pd_\Ksm  \Ham_{\Jtot_\Ksm } = 0,
	\qquad
	\pd_\Tsm \Jtot_\Ksm  + \pd_\Ksm  \Ham_\Stot = 0.
\end{equation}

This system constitutes the reversible part of the base model of heat
conduction. Before introducing the irreversible part, we note the important
property of the base model: the Hamiltonian density is conserved. Indeed,
multiplying \eqref{eq:EL5}$_1$ by $\Ham_\Stot$ and \eqref{eq:EL5}$_2$ by
$\Ham_{\Jtot_\Ksm }$ and summing the results up, we obtain:
\begin{equation}\label{eq:cons.Ham}
	\pd_\Tsm \Ham + \pd_\Ksm  \left( \Ham_\Stot \Ham_{\Jtot_\Ksm } \right) \equiv
	\Ham_\Stot\left\{ \pd_\Tsm \Stot + \pd_\Ksm  \Ham_{\Jtot_\Ksm } \right\} +
	\Ham_{\Jtot_\Ksm }\left\{ \pd_\Tsm \Jtot_\Ksm  + \pd_\Ksm  \Ham_\Stot \right\} = 0,
\end{equation}
where one can identify the spatial vector field
\begin{equation}\label{eq:heat.flux}
	q^\Ksm  := \Ham_\Stot \Ham_{\Jtot_\Ksm } = \Ttot \Ham_{\Jtot_\Ksm },
	\quad
	\text{or}
	\quad
	\qvec := \Ttot \Ham_\Jtotvec = \Ttot \JDtotvec
\end{equation}
which is nothing else than the heat flux, and where $\JDtotvec := \Ham_{\Jtotvec}$.

\begin{remark}
	System \eqref{eq:EL5} can be seen as a Hamiltonian system with the
	underlying Poisson structure. Indeed, the field equations are generated by
	the following Poisson bracket, see \cite{SHTC_GENERIC,PKG_Book2018,PPK2020}
	for details:
	\begin{equation}\label{eq:PB}
		\{F,G\} = \int \left( 
			G_\Stot \pd_\Ksm F_{\Jtot_\Ksm } 
			-
			F_\Stot \pd_\Ksm G_{\Jtot_\Ksm } 
		\right) \rmd^3 \bm{X},
	\end{equation}
	where $F$ and $G$ are arbitrary functionals of $\Stot$ and $\Jtot_\Ksm $,
	and the subscripts denote functional derivatives, e.g. $F_\Stot :=
	\frac{\delta F}{\delta \Stot}$. Note that the bracket \eqref{eq:PB} is not
	canonical. On the other hand, the canonical Poisson bracket in terms of the
	original potential $\varphi$ and the temperature $\Ttot$ reads
	\cite{Simo1988,PPK2020}
	\begin{equation}\label{eq:PB.varphi}
		\{F,G\}
		= \int \left( 
			G_{\varphi} F_{\Ttot} \right) \rmd^3 \bm{X},
			- 
			F_{\varphi} G_{\Ttot}
	\end{equation}
	which generates the second-order equation \eqref{eq:EL2} for
	$\varphi$.
\end{remark}

\subsection{Dissipation in the base model}\label{sec:base.diss}

To preserve the Hamiltonian structure \eqref{eq:EL5} of the reversible part of
the base model, the dissipation can be introduced only as lower-order terms.
For a system of first-order PDEs, the lower-order terms are algebraic source
terms:
\begin{equation}\label{eq:source}
	\pd_\Tsm \Stot + \pd_\Ksm  \Ham_{\Jtot_\Ksm } = \mathscr{S}_\Osm,
	\qquad
	\pd_\Tsm \Jtot_\Ksm  + \pd_\Ksm  \Ham_\Stot = \mathscr{S}_\Ksm ,
\end{equation}
where $\mathscr{S}_\Osm$ and $\mathscr{S}_\Ksm $ are algebraic functions of
$\Stot$ and $\Jtot_\Ksm $. They must be chosen in such a way that the first
(equation \eqref{eq:cons.Ham}) and the second law of thermodynamics are
satisfied, i.e. we have a system to be solved:
\begin{equation}
	\mathscr{S}_\Osm \geq 0,
	\qquad
	\mathscr{S}_\Osm \Ham_\Stot + \mathscr{S}_\Ksm  \Ham_{\Jtot_\Ksm } \equiv 0.
\end{equation}

From the second condition, we can express $\mathscr{S}_\Osm$ in terms of
$\mathscr{S}_\Ksm $:
\begin{equation}
	\mathscr{S}_\Osm = -\frac{1}{\Ham_\Stot} \mathscr{S}_\Ksm  \Ham_{\Jtot_\Ksm },
\end{equation}
while, we can choose the source term $\mathscr{S}_\Ksm $ in the following form:
\begin{equation}\label{eq:S_K }
	\mathscr{S}_\Ksm  = -M_{\Ksm\Lsm} \Ham_{\Jtot_L},
\end{equation}
with a symmetric positive semi-definite matrix $\bm{M}^\trans = \bm{M}\geq 0$,
$\bm{M} = \left\{ M_{\Lsm\Ksm} \right\}$ of relaxation rates. Sometimes, the choice
\eqref{eq:S_K } is called the \textit{gradient dynamics}, e.g. see
\cite{GrmelaOttingerI,GrmelaOttingerII,PKG_Book2018,SHTC_GENERIC} because it is
expressed in terms of the gradients of the Hamiltonian (thermodynamic forces).
This choice guarantees that the second law of thermodynamics is satisfied, i.e.
the entropy production is non-negative:
\begin{equation}\label{eq:SO}
	\mathscr{S}_\Osm 
	= 
	\frac{1}{\Ham_\Stot} M_{\Ksm\Lsm} \Ham_{\Jtot_L} \Ham_{\Jtot_\Ksm } \geq 0.
\end{equation}

To recover the classical Fourier law in the near-equilibrium settings, it is
sufficient to choose the matrix $\bm{M} = \text{diag}(\eps^{-1},
\eps^{-1}, \eps^{-1})$, with $\eps = \Ch^2 \tau$, and $\tau
> 0$ being the relaxation time, and the Hamiltonian density in the following
form:
\begin{equation}
	\Ham = \varepsilon(s) + \frac{1}{2} \Ch^2 \Jtot_\Ksm  \Jtot^\Ksm ,
\end{equation}
where $\varepsilon(s)$ is the internal energy density, and $\Ch$ is a
coefficient characterizing the speed of propagation of thermal perturbations.
Then, in the leading order, when $\tau \to 0$, one has:
\begin{equation}
	\Jtot_\Ksm  \approx-\tau\pd_\Ksm \Ttot
\end{equation}
and the heat flux $q^\Ksm $ in \eqref{eq:cons.Ham} reads:
\begin{equation}
	q^\Ksm = \Ttot \Ham_{\Jtot_\Ksm } = \Ch^2 \Ttot J^K
	\approx -\kappa \pd_\Ksm  \Ttot,
	\qquad
	\kappa := \tau \Ch^2 \Ttot,
\end{equation}
where $\kappa$ is the heat conductivity coefficient. In this limit, the heat
flux $q^\Ksm$ is approaching the negative temperature gradient $-\kappa\pd_\Ksm
\Ttot$, which is the classical Fourier law.

\subsection{Geometric aspects of the base model}\label{ssec:base.discussion}

Here, we briefly discuss basic geometric aspects of the base model, which will
be used as a bridge to the geometric formulation of the extended two-temperature
model in the next section. 

When we introduced the reversible part of the base model, we implied that the
four-vector field $\Jtot_\Asm$ is a gradient of the thermal displacement
potential $\varphi$, i.e. $\Jtot_\Asm = \varphi_{,\Asm} = \pd_\Asm \varphi$, and
hence the integrability condition \eqref{eq:EL3}$_2$ holds:
\begin{equation}
	\pd_\Asm \Jtot_\Bsm - \pd_\Bsm \Jtot_\Asm = 0.
\end{equation}
However, later, when we introduced the dissipative part of the base model, we
did introduce the source term $\mathscr{S}_\Ksm $ in \eqref{eq:source} which
violates the integrability condition, i.e. the antisymmetric tensor field
$\F{\Asm\Bsm}$ is no longer zero
\begin{equation}\label{eq:F.base}
	\F{\Asm\Bsm} := \pd_\Asm \Jtot_\Bsm - \pd_\Bsm \Jtot_\Asm \neq 0,
\end{equation}
and hence in the presence of dissipation, the thermal impulse $\Jtot_\Asm$
departs from being $\varphi_{,\Asm}$, the gradient of the potential $\varphi$.

In the language of differential forms, the thermal impulse $\Jtot_\Asm$ is a
1-form $\Jtotvec = \Jtot_\Asm \rmd X^\Asm $, and the antisymmetric tensor $\F{}
= \frac{1}{2}\F{\Asm\Bsm}\rmd X^\Asm \wedge \rmd X^\Bsm$ is a 2-form, which is
simply the exterior derivative of $\Jtotvec$, i.e. $\F{} = \rmd \Jtotvec$. Once
the dissipation is present, the 1-form $\Jtotvec$ is no longer exact (or
\emph{non-holonomic}), i.e. $\Jtotvec \neq \rmd \varphi$, see also
Appendix~\ref{app:base.forms} for more detailed interpretation of the base model
in the language of differential forms. In the extension of the base model
discussed in the next section, we will use the non-holonomy conditions
\eqref{eq:F.base} as new degrees of freedom to account for the non-equilibrium
and non-local effects in the heat conduction process by following the
conventional logic that the higher gradients of the state variables can be used
to represent non-resolved microscales.

Moreover, \eqref{eq:F.base} resembles the kinematics of a gauge theory where
$\varphi$ can be seen as a section of a principal bundle, $\Jtot_\Asm$ can be
interpreted as the affine connection, and $\F{\Asm\Bsm}$ is the associated
curvature tensor, or also called the field strength. Therefore, the presence of
dissipation in the base model can be interpreted as the production of the
curvature of the connection $\Jtot_\Asm$. However, the base model by no means is
a gauge theory because it deliberately does not have any local symmetry, i.e.
the Lagrangian density $\Lambda(\pd_\Asm\varphi)$ admits only global symmetries
of the form:
\begin{equation}
	\varphi(\XX) \to \varphi(\XX) + \chi,
	\qquad
	\chi = \text{const}.
\end{equation}
We note that any local symmetry would modify $J_\Asm$ and make it unobservable
(or non-measurable) like the electromagnetic potential in electrodynamics, which
contradicts the interpretation of the thermal impulse and temperature.
Nevertheless, the extended two-temperature model discussed in the rest of the
paper will have a remnant of a local gauge symmetry, which will be discussed in
Section~\ref{ssec:gauge}, and hence, a group theoretic interpretation of the
extended model will be possible in a certain sense.

\section{Non-equilibrium two-temperature model for heat conduction}\label{sec:ext.model}

In this section, we extend the base model of heat conduction to include more of
non-equilibrium effects, and provide its geometric formulation in the setting of
exterior calculus of differential forms.

\subsection{Geometric set-up of the reversible case}\label{ssec:geom.setup}

In the \emph{reversible} base model, the key field is the thermal impulse
scalar-valued $1$-form $\Jtotvec = \rmd \varphi = \Jtot_\Asm \rmd X^\Asm $,
which is exact (holonomic) $\rmd \Jtotvec = \rmd^2 \varphi = 0$. In the extended
model, we additionally introduce another 1-form $\Jmicrovec = \Jmicro_\Bsm \rmd
X^\Bsm $ which however is not implied to be holonomic even in the reversible
limit, and which represents the small-scale non-equilibrium effects in the
system (it will be the vector potential for the thermal vorticity defined
below). This construct follows a two-scale representation of a physical process
(e.g. Large Eddy Simulation in turbulence modeling). In such an approach, we
shall treat $\Jtotvec = \rmd \varphi$ as the total thermal impulse, which can be
split into a large-scale (observable/measurable) part $\Jlargevec$ and a
small-scale (fluctuating/non-observable) part $\Jmicrovec$, i.e. we have a
relationship between the three 1-forms:
\begin{equation}\label{eq:J.oneform}
	\Jlargevec = \Jtotvec - \Jmicrovec = \rmd \varphi - \Jmicrovec,
	\qquad
	\text{or}
	\qquad
	\Jtotvec = \Jlargevec + \Jmicrovec.
\end{equation}
From this decomposition, we have that $\rmd\Jlargevec =-\rmd\Jmicrovec \neq 0$,
and it is implied that the non-holonomy is a measure of the non-resolved
small-scale effects in the system. Once the small scales are completely
resolved, the non-holonomy should vanish, and one has $\Jmicrovec = 0$ and
$\Jlargevec = \Jtotvec = \rmd \varphi$, $\rmd \Jlargevec = 0$.

The temperature variables are
\begin{equation}
	\Ttot =-\Jtot_\Osm = -\pd_\Tsm\varphi,
	\qquad
	\Tmicro =-\Jmicro_\Osm,
	\qquad
	\Tlarge =-\Jlarge_\Osm = \Ttot - \Tmicro,
\end{equation}
with $\Ttot \geq 0$ being the total temperature, $\Tlarge \geq 0$ being the
large scale or measurable temperature, and $\Tmicro$ being the small-scale (or
fluctuating) temperature which has no definite sign and emerges as a
non-holonomy in time of $\Jmicrovec$.  Due to the presence of two
\emph{independent} temperature fields, we call the extended model  developed in
this section a \emph{two-temperature model}.

Furthermore, to account for non-local effects (in space and time) due to the
presence of non-resolved small-scales, we introduce the \emph{thermal vorticity}
2-form (or the \emph{field strength} in gauge-theoretic language)
\begin{equation}\label{eq:F.oneform}
	\F{} := \rmd\Jmicrovec,
	\qquad
	\F{\Asm\Bsm} = \pd_\Asm \Jmicro_\Bsm-\pd_\Bsm \Jmicro_\Asm,
	\qquad
	\rmd\F{} = \rmd^2\Jmicrovec = 0 ,
\end{equation}
which is a measure of the non-holonomy of the small-scale thermal impulse
$\Jmicrovec$. Note that, due to $\Jtotvec = \rmd \varphi$, we have that $\F{} =
\rmd \Jmicrovec =-\rmd \Jlargevec$. As we will see in Section\,\ref{sec:GK2},
the non-holonomy in time will be responsible for emergence of the longitudinal
nonlocal mode in the Guyer-Krumhansl heat conduction, while the non-holonomy in
space will be responsible for emergence of the transverse nonlocal modes.

We note that the presented geometric set up is similar to the two-scale
geometric approach used in the modeling of microstructured solids, e.g. see
\cite{NguyenLeMarrec2022,Crespo2025a,peshkov2026a,neff2014a,Madeo2015a}. In such
an approach, macroscopic and microscopic kinematic fields are attached to
the same matter-time manifold, and the \emph{incompatibility} between them
(which is due to the difference in time and length scales) is described by
geometric objects such as a connection and its torsion or curvature. In the
present context, $\varphi$ plays the role similar to that of the map between the
reference and current configurations of the medium, and $\Jmicrovec$ plays the
role similar to the non-compatible micro distortion field. Its non-holonomy
$\F{} = \rmd\Jmicrovec\neq0$ is the analogue of the geometric incompatibility
(defect density) in microstructured solids, e.g.
\cite{Clayton2005,Steinmann2015,peshkov2026a}.

Before we proceed with the derivation of the model we briefly discuss the
motivation for the choice of $\F{}$ as well as some related aspects in the
remainder of this section. Indeed, in principle, a continuum mechanics theory
could exploit the full gradient $\pd_\Asm\Jmicro_\Bsm$ (16 components), e.g. see
higher-gradient elasticity theories
\cite{Mindlin1964,Eringen1964a,DellIsola2017}, and eventually $\F{}$ is a
\emph{modeling choice}, and it is important to discuss our reasoning for it.

\subsubsection{Why do we use
$\mathbb{F}=\mathrm{d}$\textit{\boldmath{A}}?}\label{rem:whyF} 

The symmetric part of $\pd_\Asm\Jmicro_\Bsm$ is not a tensor in general, and to
transform as a tensor it requires a connection to be fixed and ordinary
derivatives to be replaced by the covariant derivatives. Therefore, a Lagrangian
depending on the full gradient (via its invariants) requires additional
geometric structure, which are essentially new degrees of freedom. On the other
hand, the exterior derivative, e.g. $\F{} = \rmd \Jmicrovec$, is defined on any
smooth manifold without a metric or an affine connection. It commutes with the
pull-back operations, and hence $\F{}$ is a tensor under arbitrary changes of
coordinates. 

Furthermore, for $\F{}=\rmd\Jmicrovec$ the integrability condition is the single
identity $\rmd\F{}=0$ (or equivalently, $\rmd\FStar{}=0$, in 4D). After the
separation of time and space ($3+1$ split) discussed later, it gives exactly one
evolution equation (for $\BFvec$ in our case). This is what produces the
Hamiltonian-like SHTC pairing \cite{SHTC_GENERIC} of a dynamical equation
(Euler-Lagrange) and a kinematic equation ($\rmd^2=0$), which can be seen in
\eqref{eq:EL3} and \eqref{eq:EL5}, as well as in the extended model
\eqref{eq:3D.ext} below. For the full gradient $\pd_\Asm\Jmicro_\Bsm$, we were
not able to find so far how to organize the integrability conditions
$\pd_{\Asm}\pd_{\Csm}\Jmicro_\Bsm - \pd_{\Csm}\pd_{\Asm}\Jmicro_\Bsm = 0$ into
such a Hamiltonian-like structure.

Finally, we note that $\F{}$ is invariant under
$\Jmicrovec\to\Jmicrovec + \rmd\chi$, whereas the full gradient is not. This is a
necessary condition if one wants to develop a gauge-theoretic interpretation of
the theory discussed in Section~\ref{ssec:gauge}.

\subsubsection{Generalization to non-torsion free manifolds}\label{rem:torsion}

Our more general interest is to geometrize the continuum mechanics and not only
heat conduction. In particular, in our previous work
\cite{PRD2019,peshkov2026a}, we developed a unified continuum mechanics theory
that treats fluids and solids as manifolds with non-zero torsion (and
potentially with non-zero curvature). It is therefore interesting to remark that
in such a geometric framework, once one starts coupling the heat conduction with
the matter sector, the coupling of the non-equilibrium fields such as thermal
vorticity 2-form $\F{}$ and the torsion 2-from $\Tors{\Asm}{}{}$ already starts
to occur at the geometric (or kinematic) level.

Indeed, in this case, the coordinate basis $\rmd X^\Asm$ has to be replaced by a
non-coordinate basis $\bm{h}^\Asm $, which is non-holonomic in general and has a
non-zero torsion $\Tors{\Asm}{}{}=\rmd \bm{h}^\Asm  =
\frac{1}{2}\Tors{\Asm}{\Bsm}{\Csm}\bm{h}^\Bsm \wedge \bm{h}^\Csm $, and
hence taking the exterior derivative of $\Jmicrovec = \Jmicro_\Bsm \bm{h}^\Bsm $
gives
\begin{equation}
	\F{} 
	= \rmd \Jmicrovec 
	= \rmd \Jmicro_\Bsm\wedge \bm{h}^\Bsm  + \Jmicro_\Bsm \rmd \bm{h}^\Bsm,
\end{equation}
we can see that the second term in the right hand side does not disappear in
general, which is exactly the torsion 2-form that contributes in the definition
of the thermal vorticity. In component form, this reads
\begin{equation}
	\F{\Asm\Bsm} 
	= 
	\pd_\Asm \Jmicro_\Bsm - \pd_\Bsm \Jmicro_\Asm
	+ \Tors{\Csm}{\Asm}{\Bsm}\Jmicro_\Csm.
\end{equation}
Of course, we still have the identity $\rmd \F{} = \rmd^2 \Jmicrovec = 0$.
Therefore, if we believe that macroscopic models (e.g. continuum mechanics
models) should obey a geometric structure, then we can see that certain fields
already have to be coupled at the kinematic level (beyond constitutive
relations), which otherwise might be overlooked.

\subsection{Variational formulation}\label{sec:ext.variational}

We assume that the action integral is a functional of two potentials: the
thermal displacement and microscopic thermal impulse $\Jmicro_\Bsm$:
\begin{equation}
	S[\varphi,\Jmicrovec] = \int
	\Lambda(\varphi_{,\Asm},\Jmicro_\Bsm,\pd_\Asm \Jmicro_\Bsm)
	\rmd^4 \XX,
\end{equation}
and it is implied that the Lagrangian density $\Lambda$ may depend on $\varphi$
and $\Jmicro_\Bsm$ and their first derivatives $\pd_\Asm\varphi$ and
$\Jmicro_{\Bsm,\Asm}$. However, in the following, we shall consider the case
where we dropped out the dependence of $\Lambda$ on $\varphi$ (but not on
$\Jmicro_\Bsm$) from the very beginning, i.e. $\Lambda_\varphi = 0$.

The Euler-Lagrange equations for the variation $\delta\varphi$ read (recall that
$\Jtot_\Asm = \varphi_{,\Asm}$):
\begin{equation}\label{eq:EL.varphi}
	\frac{\delta S}{\delta \varphi} = \Lambda_\varphi
	- \pd_\Asm \Lambda_{\Jtot_\Asm} =- \pd_\Asm \Lambda_{\Jtot_\Asm} = 0,
\end{equation}
and for the variation $\delta \Jmicrovec$:
\begin{equation}\label{eq:EL.ext}
	\frac{\delta S}{\delta \Jmicro_\Bsm} =
	\Lambda_{\Jmicro_\Bsm}
	- \partial_\Asm \Lambda_{\Jmicro_{\Bsm,\Asm}} = 0,
\end{equation}
where $\Jmicro_{\Bsm,\Asm} = \pd_\Asm \Jmicro_\Bsm$.

As it was motivated in Section~\ref{rem:whyF}, we shall work not with the full
gradient $\pd_\Asm\Jmicro_\Bsm$ but with its anti-symmetric part, the thermal
vorticity $\F{} = \rmd \Jmicrovec$. In particular, we prefer to work with the
Hodge dual $\FStar{}$ of the thermal vorticity $\F{}$. Since the dimension of
the matter-time manifold is four $n=4$, the Hodge dual of a 2-form is again a
2-form ($n-2$), or more precisely, ${}^\star\FStar{} =-\F{}$. In this special
case, we have the identity
\begin{equation}\label{eq:dFStar}
	\delta(\FStar{}) = 0,
\end{equation}
where $\delta$ is the codifferential operator, which is the adjoint of the
exterior derivative $\rmd$ with respect to the Hodge star operator. In the
coordinate form, identity \eqref{eq:dFStar} is a divergence identity that reads:
\begin{equation}
	\FStar{\Asm\Bsm} := \frac{1}{2} \veps^{\Asm\Bsm\Csm\Dsm} \F{\Csm\Dsm} = \veps^{\Asm\Bsm\Csm\Dsm} \pd_\Csm \Jmicro_\Dsm ,
	\qquad
	\pd_\Bsm \FStar{\Asm\Bsm} 
	= \veps^{\Asm\Bsm\Csm\Dsm} \pd_\Bsm \pd_\Csm \Jmicro_\Dsm 
	\equiv 0.
\end{equation}

Therefore, we imply the following re-parametrization of the Lagrangian density
($\Lambda \to L$):
\begin{equation}\label{eq:Lambda.L}
	\Lambda(\Jtot_\Asm,\Jmicro_\Bsm,\pd_\Asm\Jmicro_\Bsm) 
	= 
	L(\Jtot_\Asm,\Jmicro_\Bsm,\FStar{\Asm\Bsm}),
\end{equation}
which results in the following relations between the derivatives:
\begin{equation}
	\Lambda_{\Jtot_\Asm} = L_{\Jtot_\Asm},
	\qquad
	\Lambda_{\Jmicro_{\Bsm,\Asm}} = \veps^{\Asm\Bsm\Csm\Dsm} L_{\FStar{\Csm\Dsm}}
	:= \Hexci{\Asm\Bsm},
\end{equation}
where the notation $\Hexci{\Asm\Bsm}$ is introduced for the excitation tensor.
After reparametrization \eqref{eq:Lambda.L}, the Euler-Lagrange equation
\eqref{eq:EL.ext} becomes:
\begin{equation}\label{eq:EL.Jmicro}
	\Lambda_{\Jmicro_{\Bsm}} - \pd_\Asm \Lambda_{\Jmicro_{\Bsm,\Asm}}
	= L_{\Jmicro_{\Bsm}}
	- \pd_\Asm \left( \veps^{\Asm\Bsm\Csm\Dsm} L_{\FStar{\Csm\Dsm}} \right)
	= L_{\Jmicro_{\Bsm}} - \pd_\Asm \Hexci{\Asm\Bsm}
	= 0.
\end{equation}

The two Euler-Lagrange equations \eqref{eq:EL.varphi} and \eqref{eq:EL.Jmicro}
can be accompanied by the identity $\rmd^2\varphi = \rmd \Jtotvec = 0$ and the
definition $\F{} = \rmd \Jmicrovec$, as well as by two Bianchi-type identities
$\pd_\Bsm \FStar{\Asm\Bsm} = 0$  and $\pd_\Bsm L_{\Jmicro_\Bsm} = 0$ (the latter
can be obtained from the Euler-Lagrange equation \eqref{eq:EL.Jmicro} after
applying the divergence operator and using the fact that the excitation tensor
$\Hexci{\Asm\Bsm}$ is antisymmetric) to form the final system of field equations
in the phase space $(\Jtot_\Asm,\Jmicro_\Bsm,\FStar{\Asm\Bsm})$:
\begin{subequations}\label{eq:4D.ext}
	\begin{align}
		\pd_\Asm L_{\Jtot_{\Asm}} &= 0,\label{eq:4D.ext.Slarge}\\[2mm]
		\pd_\Asm \Jtot_{\Bsm} - \pd_\Bsm \Jtot_{\Asm} &= 0\label{eq:4D.ext.Jobs},\\[2mm]
		\pd_\Bsm L_{\Jmicro_{\Bsm}} &= 0,\label{eq:4D.ext.s}\\[2mm]
		\pd_\Asm \Jmicro_{\Bsm} - \pd_\Bsm \Jmicro_{\Asm} &= \F{\Asm\Bsm}
		\label{eq:4D.ext.Jmicro},\\[2mm]
		\pd_\Asm \Hexci{\Bsm\Asm} &= -L_{\Jmicro_{\Bsm}},\label{eq:4D.ext.D}\\[2mm]
		\pd_\Bsm \FStar{\Asm\Bsm} &= 0.\label{eq:4D.ext.B}
	\end{align}
\end{subequations}

\subsection{3D version of the two-temperature model}\label{sec:ext.3D}

In this section, we rewrite system \eqref{eq:4D.ext} in the 3D notations
associated with the observer co-moving with matter (Lagrangian observer). The
four-velocity of such an observer with respect to the material coordinates
$X^\Asm $ is trivial: $U^\Asm  = \frac{\pd X^\Asm }{\pd X^\Osm} = \delta^\Asm
_{\Osm} = (1,0,0,0)$.

Following almost exactly the steps from \cite[Sec.VII--VIII]{Peshkov2025}, we
introduce the following projections of the antisymmetric tensor $\F{\Asm\Bsm}$
and of its Hodge dual $\FStar{\Asm\Bsm}$:
\begin{equation}
	\EF{\Asm} := \F{\Asm\Bsm} U^\Bsm ,
	\qquad
	\BF{\Asm} := \FStar{\Asm\Bsm} U_\Bsm,
\end{equation}
so that
\begin{equation}
	\F{\Asm\Bsm} = U_\Asm \EF{\Bsm} - U_\Bsm \EF{\Asm}
	- \veps_{\Asm\Bsm\Csm\Dsm} U^\Csm \BF{\Dsm},
	\qquad
	\FStar{\Asm\Bsm} = U^\Asm  \BF{\Bsm} - U^\Bsm  \BF{\Asm}
	+ \veps^{\Asm\Bsm\Csm\Dsm} U_\Csm \EF{\Dsm},
\end{equation}
where $U_\Asm = G_{\Asm\Bsm} U^\Bsm = (-1,0,0,0)$. Note that in such a frame the
time components of $\EF{\Asm}$ and $\BF{\Asm}$ vanish, i.e. $\EF{\Osm} = 0$ and
$\BF{\Osm} = 0$.

We then again reparametrize the Lagrangian density $L$ from \eqref{eq:Lambda.L}
in terms of $\EF{\Asm}$ and $\BF{\Asm}$ as
\begin{equation}
	L(\Jtot_\Asm, \Jmicro_\Bsm, \FStar{\Asm\Bsm})
	= \mathcal{L}(\Jtot_\Asm, \Jmicro_\Bsm,\EF{\Asm},\BF{\Asm}),
\end{equation}
and perform a partial Legendre transformation (only with respect to the time
components of $\Jtot_\Asm$, $\Jmicro_\Bsm$, and $\F{\Asm\Bsm}$) to introduce the
Hamiltonian density
$\mathcal{H}(\Stot,\Jtot_\Ksm ,\Smicro,\Jmicro_\Ksm ,\DF{\Ksm},\BF{\Ksm})$:
\begin{equation}
	\mathcal{H}
	:= \Ttot \mathcal{L}_\Ttot 
	+ \Tmicro \mathcal{L}_\Tmicro 
	+ \EF{\Ksm} \mathcal{L}_{\EF{\Ksm}} 
	- \mathcal{L},
\end{equation}
with the following relations between the dual variables and the original
variables:
\begin{subequations}
	\begin{align}
		\Stot := \mathcal{L}_{\Ttot},
		\qquad
		&\Ttot = \Ham_\Stot, \\[2mm]
		\Smicro := \mathcal{L}_{\Tmicro},
		\qquad
		&\Tmicro = \Ham_\Smicro, \\[2mm]
		\DF{\Ksm} := \mathcal{L}_{\EF{\Ksm}},
		\qquad
		&\EF{\Ksm} = \Ham_{\DF{\Ksm}},
	\end{align}
\end{subequations}
while for the spatial components of the thermal impulse and thermal vorticity, we have
\begin{equation}
	-\mathcal{L}_{\Jtot_\Ksm } = \Ham_{\Jtot_\Ksm },
	\quad
	-\mathcal{L}_{\Jmicro_\Ksm } = \Ham_{\Jmicro_\Ksm },
	\quad
	-\mathcal{L}_{\BF{\Ksm}} = \Ham_{\BF{\Ksm}}.
\end{equation}
The newly introduced variables $\Stot = \mathcal{L}_{\Ttot}$ and $\Smicro =
\mathcal{L}_{\Tmicro}$ are the \emph{total entropy} and the \emph{fluctuating
entropy}.

After these steps, system \eqref{eq:4D.ext} can be written in the canonical SHTC
form \cite{GodRom1996,SHTC_GENERIC}:
\begin{subequations}\label{eq:3D.ext}
	\begin{align}
		\pd_\Tsm \Stot + \pd_\Ksm  \Ham_{\Jtot_\Ksm} & = 0, \label{eq:3D.ext.Sobs}\\[2mm]
		\pd_\Tsm \Jtot_\Ksm  + \pd_\Ksm  \Ham_\Stot  & = 0, \label{eq:3D.ext.Jobs}\\[2mm]
		\pd_\Tsm \Smicro + \pd_\Ksm  \Ham_{\Jmicro_\Ksm}  & = 0, \label{eq:3D.ext.Smicro}\\[2mm]
		\pd_\Tsm \Jmicro_\Ksm  + \pd_\Ksm  \Ham_\Smicro & = - \Ham_{\DF{\Ksm}}, \label{eq:3D.ext.Jmicro}\\[2mm]
		\pd_\Tsm \DF{\Ksm}
		- \pd_\Lsm \left(\veps^{\Ksm\Lsm\Ism} \Ham_{\BF{\Ism}} \right)
		& = + \Ham_{\Jmicro_\Ksm }, \label{eq:3D.ext.D}\\[2mm]
		\pd_\Tsm \BF{\Ksm}
		+ \pd_\Lsm \left(\veps^{\Ksm\Lsm\Ism} \Ham_{\DF{\Ism}} \right)
		& = 0.\label{eq:3D.ext.B}
	\end{align}
\end{subequations}

Note that the $\Osm$-th equations of \eqref{eq:4D.ext.D}, \eqref{eq:4D.ext.B} and
\eqref{eq:4D.ext.Jmicro} reduce to the following involution constraints:
\begin{equation}\label{eq:involutions}
	\pd_\Ksm  \DF{\Ksm} =-\Smicro,
	\qquad
	\pd_\Ksm  \BF{\Ksm} = 0,
	\qquad
	\veps^{\Ksm\Ism\Lsm} \pd_\Ism \Jmicro_\Lsm = \BF{\Ksm}.
\end{equation}
or, in vector calculus notations:
\begin{equation}\label{eq:involutions0}
	\nabla\cdot\DFvec = -\Smicro,
	\qquad
	\nabla\cdot\BFvec = 0,
	\qquad
	\nabla\times\Jmicrovec = \BFvec.
\end{equation}

System \eqref{eq:3D.ext} admits the following energy conservation law (it can be
obtained by multiplying every equation by the corresponding dual variable and
summing them up):
\begin{subequations}\label{eq:consH.ext}
	\begin{align}
		\pd_\Tsm \Ham
		+ \pd_\Ksm  \left( \Ham_\Stot \Ham_{\Jtot_\Ksm }
		+ \Ham_\Smicro \Ham_{\Jmicro_\Ksm }
		+ \veps^{\Ksm\Lsm\Ism} \Ham_{\DF{\Lsm}} \Ham_{\BF{\Ism}} \right)
		&= {\color{blue}\Ham_\Stot} \left\{ \pd_\Tsm \Stot + \pd_\Ksm  \Ham_{\Jtot_\Ksm } \right\}	\label{eq:cons.E.ext}\\[2mm]
		&+ {\color{blue}\Ham_{\Jtot_\Ksm }} \left\{ \pd_\Tsm \Jtot_\Ksm  + \pd_\Ksm  \Ham_\Stot \right\} \\[2mm]
		&+ {\color{blue}\Ham_{\Smicro}} \left\{ \pd_\Tsm \Smicro + \pd_\Ksm  \Ham_{\Jmicro_\Ksm } \right\} \\[2mm]
		&+ {\color{blue}\Ham_{\Jmicro_\Ksm }} \left\{ \pd_\Tsm \Jmicro_\Ksm  + \pd_\Ksm  \Ham_\Smicro + \Ham_{\DF{\Ksm}} \right\} \\[2mm]
		&+ {\color{blue}\Ham_{\DF{\Ksm}}} \left\{ \pd_\Tsm \DF{\Ksm}
		- \pd_L \left(\veps^{\Ksm\Lsm\Ism} \Ham_{\BF{\Ism}} \right)
		- \Ham_{\Jmicro_\Ksm } \right\} \\[2mm]
		&+ {\color{blue}\Ham_{\BF{\Ksm}}} \left\{ \pd_\Tsm \BF{\Ksm}
		+ \pd_\Lsm \left(\veps^{\Ksm\Lsm\Ism} \Ham_{\DF{\Ism}} \right)
		\right\} = 0.
	\end{align}
\end{subequations}

We note that the source terms in \eqref{eq:3D.ext.Jmicro} and
\eqref{eq:3D.ext.D} are summed up to 0 in the summation
\eqref{eq:consH.ext}:
\begin{equation}
	{\color{blue}\Ham_{\Jmicro_\Ksm }} \Ham_{\DF{\Ksm}}
	- {\color{blue}\Ham_{\DF{\Ksm}}} \Ham_{\Jmicro_\Ksm }
	\equiv 0,
\end{equation}
which is a consequence of their skew-symmetric structure. In other words, these
source terms do not have to be counterbalanced by any additional source terms
in the entropy balance law \eqref{eq:3D.ext.Sobs} to satisfy the energy
conservation law \eqref{eq:consH.ext} as it was done in the base model, see
equations \eqref{eq:source}--\eqref{eq:SO}. This means that source terms in
\eqref{eq:3D.ext.Jmicro} and \eqref{eq:3D.ext.D} are of reversible nature, which
is not surprising as they are coming from the variational principle.

The energy flux \eqref{eq:cons.E.ext} suggests that the heat flux vector in the
extended two-temperature model is given by
\begin{equation}\label{eq:q.ext}
	q^\Ksm = \Ham_\Stot \Ham_{\Jtot_\Ksm }
		+ \Ham_\Smicro \Ham_{\Jmicro_\Ksm }
		+ \veps^{\Ksm\Lsm\Ism} \Ham_{\DF{\Lsm}} \Ham_{\BF{\Ism}},
	\quad
	\text{or}
	\quad
	\qvec = \Ttot \JDtotvec + \Tmicro \JDmicrovec + \EFvec \times \HFvec,
\end{equation}
where $\JDtotvec := \Ham_{\Jtotvec}$, $\JDmicrovec := \Ham_{\Jmicrovec}$, and
$\HFvec := \Ham_{\BFvec}$, i.e. it has an additional (if compared to the base
model \eqref{eq:heat.flux}) non-equilibrium contribution from the fluctuating
heat flux $\Tmicro\JDmicrovec$ and a non-local effect due to the thermal
Poynting vector $\EFvec \times \HFvec$.

Remark that it is usual that extended models of heat conduction introduce the
concept of the \emph{extra entropy flux} (extra with respect to $\qvec/\Ttot$),
e.g. see \cite{giorgi2026}. In this sense, the entropy flux $\JDtotvec =
\Ham_{\Jtotvec}$ of our model can be also split into the classical part
$\qvec/\Ttot$ and the extra entropy flux:
\begin{equation}
	\JDtotvec = \frac{1}{\Ttot} \qvec
	- \frac{1}{\Ttot} (\Tmicro \JDmicrovec 
	+ \EFvec\times\HFvec).
\end{equation}

\subsection{Rescaled two-temperature system}\label{ssec:rescaling.alpha} 

For the rest of the paper, we shall use the vector calculus notations, and we
rescale $\DFvec$ and $\BFvec$ fields by a constant $\alpha$:
\begin{equation}\label{eq:alpha}
	\DFvec \to \alpha^{-1} \DFvec = \tilde{\DFvec},
	\quad
	\BFvec \to \alpha \BFvec = \tilde{\BFvec},
\end{equation}
and keep others unchanged. Such a rescaling results in (recall that $\EFvec =
\Ham_{\DFvec}$ and $\HFvec = \Ham_{\BFvec}$):
\begin{equation}
	\EFvec \to \alpha     \EFvec = \tilde{\EFvec},
	\quad
	\HFvec \to \alpha^{-1}\HFvec = \tilde{\HFvec}.
\end{equation}

After this, system \eqref{eq:3D.ext} now reads (where we dropped the
tildes for simplicity):
\begin{subequations}\label{eq:3D.ext.rescaled}
	\begin{align}
		\pd_\Tsm \Stot + \nabla\cdot\Ham_{\Jtotvec} &= 0,\\[1mm]
		\pd_\Tsm \Jtotvec + \nabla\Ham_\Stot &= 0,\\[1mm]
		\pd_\Tsm \Smicro + \nabla\cdot\Ham_{\bm A} &=0,\\[1mm]
		\pd_\Tsm \bm{\Asm} + \nabla\Ham_{\Smicro} &=-\frac{1}{\alpha}\Ham_{\DFvec},\\[1mm]
		\pd_\Tsm \DFvec - \nabla\times\Ham_{\BFvec} &=
		+ \frac{1}{\alpha}\Ham_{\bm A},\\[1mm]
		\pd_\Tsm \BFvec + \nabla\times\Ham_{\DFvec} &= 0.
	\end{align}
\end{subequations}
The involutions \eqref{eq:involutions0} for the rescaled system become
\begin{equation}\label{eq:inv.rescaled}
	\nabla\cdot\DFvec=-\alpha^{-1} \Smicro,
	\qquad
	\nabla\cdot\BFvec = 0,
	\qquad
	\BFvec = \alpha \nabla\times\Jmicrovec.
\end{equation}

\subsection{Dissipation in the two-temperature model}\label{ssec:2temp.diss}

The dissipation (consistent with the total energy conservation
\eqref{eq:consH.ext}) in the two-temperature model can be introduced in the
following way:
\begin{subequations}\label{eq:3D.2temp.diss}
	\begin{align}
		\pd_\Tsm \Stot + \nabla\cdot\Ham_{\Jtotvec} &=
		  \frac{1}{\Ttot\eps}\Ham_{\Jtotvec}^2
		+ \frac{1}{\Ttot\sigma}\Ham_{\Smicro}^2
		+ \frac{1}{\Ttot\eta}\Ham_{\DFvec}^2,
		\label{eq:3D.nonhol.diss.s}\\[1mm]
		\pd_\Tsm \Jtotvec + \nabla\Ham_\Stot &=-\frac{1}{\eps}\Ham_{\Jtotvec},
		\label{eq:3D.nonhol.diss.Jobs}\\[1mm]
		\pd_\Tsm \Smicro + \nabla\cdot\Ham_{\bm A} &=-\frac{1}{\sigma}\Ham_{\Smicro},
		\label{eq:3D.nonhol.diss.s2}\\[1mm]
		\pd_\Tsm \bm{\Asm} + \nabla\Ham_{\Smicro} &=-\frac{1}{\alpha}\Ham_{\DFvec},
		\label{eq:3D.nonhol.diss.A}\\[1mm]
		\pd_\Tsm \DFvec - \nabla\times\Ham_{\BFvec} &=
		+ \frac{1}{\alpha}\Ham_{\bm A} - \frac{1}{\eta}\Ham_{\DFvec},
		\label{eq:3D.nonhol.diss.D}\\[1mm]
		\pd_\Tsm \BFvec + \nabla\times\Ham_{\DFvec} &= 0,
		\label{eq:3D.nonhol.diss.B}
	\end{align}
\end{subequations}
 where $\eps = \Ch^2\tau > 0$ is defined as in the base model, $\sigma>0$, and
 $\eta>0$ are new relaxation parameters.

 Note that the involution $\nabla\cdot\DFvec=-\alpha^{-1} s$ is not preserved by
the dissipative system, and this stationary relation becomes an evolution
equation:
\begin{equation}\label{eq:thermal.charge}
	\pd_\Tsm \rho + \nabla\cdot(\eta^{-1}\EFvec) =-\frac{1}{\alpha\sigma} \Tmicro,
	\qquad
	\rho:=\nabla\cdot\DFvec + \alpha^{-1} s,
\end{equation}
where $\rho$ can be called the \emph{free thermal charge density} in analogy
with electrodynamics.

\section{Guyer-Krumhansl-type heat conduction as a relaxation limit}
\label{sec:GK2}

In this section, we discuss a relaxation limit of the non-equilibrium
two-temperature model \eqref{eq:3D.2temp.diss} for a simple quadratic
Hamiltonian, and demonstrate that it leads to a Guyer-Krumhansl-type (GK) heat
conduction \cite{Guyer1966,EIT2010,Van2017,Zhu2017}.

Before performing the asymptotic analysis of \eqref{eq:3D.2temp.diss}, the
system has to be closed by specifying the dependence of the Hamiltonian density
$\Ham(\bm{w})$ on the state variables $\bm{w} =
\{\Stot,\Jtotvec,\Smicro,\Jmicrovec,\DFvec,\BFvec\}$. In particular, we shall
assume a simple Hamiltonian density
\begin{equation}\label{eq:H.quadratic.GK2}
	\Ham =
	\veps(\Stot)
	+ \frac{1}{2c_s} \Smicro^2
	+ \frac{1}{2}\Ch^2 \Vert\Jtotvec - \Jmicrovec\Vert^2
	+ \frac{1}{2}\ch^2 \Vert\Jmicrovec\Vert^2
	+ \frac{1}{2\mu_0} \Vert \BFvec\Vert^2
	+ \frac{1}{2\epsilon_0} \Vert \DFvec\Vert^2,
\end{equation}
which is quadratic almost in all variables, except for the total entropy
$\Stot$. Here, $\veps(\Stot)$ is the internal energy, $\Ch > 0 $, $\ch > 0 $,
$\epsilon_0>0$, and $\mu_0>0$ are some material parameters, while $c_s>0$ is the
specific heat of the fluctuating entropy.

The dual variables $\Ttot = \Ham_{\Stot}$, $\Tmicro = \Ham_{\Smicro}$,
$\JDmicrovec = \Ham_{\Jmicrovec}$, etc. are then given by (recall that
$\Jlargevec = \Jtotvec - \Jmicrovec$)
\begin{subequations}\label{eq:quadratic.duals.GK2}
	\begin{align}
		\Ttot = \Ham_{\Stot} = \veps'(\Stot),
		\quad
		& \Tmicro = \Ham_{\Smicro} = \frac{\Smicro}{c_\Smicro},
		\\
		\JDtotvec = \Ch^2\Jlargevec,
		\quad
		& \JDmicrovec =-\Ch^2\Jlargevec + \ch^2\Jmicrovec,
		\\
		\EFvec=\frac{\DFvec}{\epsilon_0},
		\quad
		&\HFvec=\frac{\BFvec}{\mu_0}.
	\end{align}
\end{subequations}

From the dimensional analysis, one can see that the parameter $\eta$ is related
to $\alpha$ and $\eps $ (defined after the eq.\eqref{eq:SO}) as $\eta = \alpha^2
\eps^{-1}$. In particular, in the rescaling procedure \eqref{eq:alpha}, $\alpha$
can be chosen in such a way that the ratio $\lambda = \alpha / \eps$ becomes
nondimensional. In particular, we have
\begin{equation}\label{eq:alpha.eta.lambda.GK2}
	\alpha = \eps\lambda,
	\qquad
	\eta = \eps\lambda^2,
\end{equation}
so that $\lambda\to0$ means $\alpha\to0$ and $\eta\to 0$ at fixed $\eps$.
Therefore, we consider expansion of the fields in powers of the non-dimensional
parameter $\lambda$. For $\sigma$ we only assume that the fluctuation
temperature $\Tmicro$ relaxes fast compared with the relaxation time scale
$\tau_1 = \tau$ of macroscopic thermal impulse $\Jtotvec$, i.e. we assume that
$\sigma c_\Smicro\ll\tau_1$. Its precise order in $\lambda$ has to be $\sigma =
\mathcal{O}(\lambda^2)$ to get the correct Guyer-Krumhansl limit.

The relaxation time $\tau_1 = \tau$ and $\eps = \Ch^2\tau_1$ introduces another
relaxation time
\begin{equation}\label{eq:tau2}
	\tau_2 := \frac{\eps}{\ch^2} = \frac{\Ch^2}{\ch^2}\,\tau_1 ,
\end{equation}
so that $\eps^{-1}\JDtotvec = \tau_1^{-1}\Jlargevec$ and $\eps^{-1}\JDmicrovec =
-\tau_1^{-1}\Jlargevec + \tau_2^{-1}\Jmicrovec$. In the analysis below, $\Ch$ and
$\ch$ (and hence $\tau_1$ and $\tau_2$) are kept fixed.

\subsection{The temperature equations}

Equation \eqref{eq:3D.nonhol.diss.s2} on the fluctuating entropy can be
re-written as
\begin{equation}\label{eq:theta.micro.PDE}
	-\sigma c_\Smicro \pd_\Tsm\Tmicro
	-\sigma \nabla\cdot \JDmicrovec = \Tmicro.
\end{equation}
Therefore, for processes slow compared with $\sigma c_\Smicro$, we have $\Tmicro
\approx -\sigma \nabla\cdot\JDmicrovec$, and hence
\begin{equation}\label{eq:theta.micro.approx}
	\nabla\Tmicro \approx -\sigma \nabla(\nabla\cdot\JDmicrovec).
\end{equation}
This relation will be responsible for the longitudinal non-local part of the
GK-type heat conduction.

The total entropy equation \eqref{eq:3D.nonhol.diss.s} has only quadratic
terms in thermodynamic forces, which are assumed to be small, and it becomes an
equation for the macroscopic temperature:
\begin{equation}
	c_\Stot\pd_\Tsm\Ttot + \Ttot_0 \nabla\cdot\JDtotvec = 0,
\end{equation}
with $c_\Stot = \Ttot_0/\veps''(\Stot_0)$ being the specific heat of the
macroscopic entropy, and $\Ttot_0 = \veps'(\Stot_0)$ being the macroscopic
temperature at the reference state.

\subsection{Formal asymptotic expansion}

We shall only need the following three PDEs from the system
\eqref{eq:3D.2temp.diss}:
\begin{align}
	\EFvec
	&= \eps\lambda\left(\pd_\Tsm\Jlargevec + \nabla\Tlarge
	+ \frac{1}{\eps}\JDtotvec \right),\label{eq:E.from.Jmacro}\\[2mm]
	\EFvec
	&= \lambda \JDmicrovec
	- \eps_0 \eps \lambda^2 \pd_\Tsm\EFvec
	+ \eps\lambda^2 \nabla\times\HFvec,\label{eq:E.from.D}\\[2mm]
	\BFvec
	&= \eps\lambda\,\nabla\times\Jmicrovec
	= \eps\lambda\nabla\times(\Jtotvec - \Jlargevec),\label{eq:B.rescaled}
\end{align}
where \eqref{eq:E.from.Jmacro} is the difference of
\eqref{eq:3D.nonhol.diss.Jobs} and \eqref{eq:3D.nonhol.diss.A}, the second
equation is \eqref{eq:3D.nonhol.diss.D}, and the third equation is
\eqref{eq:inv.rescaled}$_3$. Additionally, we will need an equation on
$\Jtotcurl := \nabla\times\Jtotvec$ :
\begin{equation}\label{eq:Omega.2T}
	\pd_\Tsm\Jtotcurl = -\frac{1}{\eps}\nabla\times\JDtotvec,
\end{equation}
which can be obtained by taking the curl of \eqref{eq:3D.nonhol.diss.Jobs}.


Equation \eqref{eq:E.from.D} states that 
\begin{equation}
	\EFvec = \lambda \JDmicrovec + \mathcal{O}(\lambda^3),
\end{equation}
which if plugged in into \eqref{eq:E.from.Jmacro} gives
\begin{equation}
	\pd_\Tsm \Jlargevec 
	+ \nabla \Tlarge 
	+ \frac{1}{\eps}\left( \JDtotvec - \JDmicrovec \right)
	= \mathcal{O}(\lambda^2).
\end{equation}
In other words, equation \eqref{eq:E.from.Jmacro} up to
$\mathcal{O}(\lambda^2)$, is our base model with $\tau_{\mathrm{eff}} = \tau/2$,
which is similar to the hyperbolic Cattaneo heat conduction. In other words, the
expansion goes in powers of $\lambda^2$.

The first non-trivial correction to the Cattaneo heat conduction appears at the
order $\mathcal{O}(\lambda^3)$. Indeed, plugging \eqref{eq:E.from.Jmacro} into
\eqref{eq:E.from.D} and dividing by $\eps\lambda$, at order
$\mathcal{O}(\lambda^3)$, we obtain
\begin{equation}\label{eq:general.GK}
	\pd_\Tsm(\Jlargevec + \eps_0 \lambda^2 \JDmicrovec)
	+ \nabla\Ttot  + \sigma \nabla(\nabla\cdot\JDmicrovec) 
	- \lambda\nabla\times\HFvec
	+ \frac{1}{\eps}(\JDtotvec - \JDmicrovec) 
	= \mathcal{O}(\lambda^3),
\end{equation}
where we have used that $\Tlarge = \Ttot - \Tmicro \approx \Ttot +
\sigma\nabla\cdot\JDmicrovec$, and also that
\begin{equation}
	\eps_0\eps\lambda^2\pd_\Tsm\left( \pd_\Tsm\Jlargevec
	+ \nabla\Tlarge
	+ \frac{1}{\eps}\JDtotvec \right) =
	 \eps_0\lambda^2\pd_\Tsm\JDmicrovec
	+ \mathcal{O}(\lambda^3).
\end{equation}

Finally, to arrive at a Guyer-Krumhansl-type equation, we need to use the
constitutive relation $\HFvec = \BFvec/\mu_0 = \eps\lambda\nabla\times(\Jtotvec
- \Jlargevec)/\mu_0$ and express thermal impulses via the heat flux $\qvec$. For
this purpose, we use the fact that in leading orders, the heat flux
\eqref{eq:q.ext} of our model linearized around the temperature $\Ttot_0$ and
vanishing fluctuating temperature $\Tmicro_0 = 0$ is given by (the Poynting
vector is quadratic and hence it was dropped out)
\begin{equation}
	\qvec = \Ttot_0 \JDtotvec = \Ttot_0 \Ch^2 \Jlargevec,
	\quad
	\text{or}
	\quad
	\Jlargevec = \frac{1}{\Ttot_0 \Ch^2}\qvec,
\end{equation}
and hence
\begin{equation}
	\JDmicrovec 
	=-\Ch^2\Jlargevec + \ch^2\Jmicrovec
	=-\frac{1}{\Ttot_0}\qvec + \ch^2\Jmicrovec.
\end{equation}

Then, replacing $\JDtotvec$, $\Jlargevec$, and $\JDmicrovec$ in
\eqref{eq:general.GK} by the above expressions, and re-arranging terms, we
obtain the following Guyer-Krumhansl-type equation for the heat flux $\qvec$:
\begin{subequations}\label{eq:GK.2T}
	\begin{equation}\label{eq:GK.2T.q}
		\pd_\Tsm \qvec 
		+ \frac{\kappa}{\tau_1'} \nabla\Ttot
		+ \frac{\ell_T^2}{\tau_1'} \nabla \times\left( \nabla\times
		\qvec - \Ttot_0 \Ch^2 \Jtotcurl \right)
		- \frac{3\ell_L^2}{\tau_1'} \nabla (\nabla\cdot\qvec)
		+ \frac{1}{\tau_1'} \qvec
		= 
		-\frac{1}{\tau_2'}\mathcal{R}(\Jmicrovec) ,
	\end{equation}
	\begin{equation}\label{eq:GK.2T.params}
		\beta_0 = 1 - \eps_0 \Ch^2 \lambda^2,
		\qquad
		\tau_1' = \frac{\beta_0\tau_1}{2},
		\qquad
		\kappa = \frac{\Ttot_0\Ch^2\tau_1}{2},
		\qquad
		\ell_T^2 = \frac{\eta\tau_1}{2\mu_0},
		\qquad
		\ell_L^2 = \frac{\sigma\Ch^2\tau_1}{6},
	\end{equation}
	and where we have a relaxation of the Cattaneo-type residual
	$\mathcal{R}(\Jmicrovec)$ of the microscopic thermal impulse:
	\begin{equation}\label{eq:GK.2T.A}
		\mathcal{R}(\Jmicrovec) = 
		\Ttot_0 \Ch^2 \left( 
			  \eps_0 \eps \lambda^2 \pd_\Tsm\Jmicrovec
			+ \eps \sigma \nabla(\nabla\cdot\Jmicrovec) 
			- \Jmicrovec
			\right),
			\qquad
			\tau_2' = \beta_0\tau_2.
		\end{equation}
\end{subequations}

Finally, assuming $\Jtotcurl(0) = 0$ in the initial data, \eqref{eq:Omega.2T}
integrates to 
\begin{equation}
	\Jtotcurl = -\frac{1}{\eps}\int_0^T\nabla\times\JDtotvec\,\rmd T' = -\frac{1}{\eps \Ttot_0}\int_0^T\nabla\times\qvec\,\rmd T',
\end{equation}
and equation \eqref{eq:GK.2T.q} can be written as
\begin{equation}\label{eq:GK.2T.q.int}
	\pd_\Tsm \qvec 
	+ \frac{\kappa}{\tau_1'} \nabla\Ttot
	+ \frac{\ell_T^2}{\tau_1'} \nabla \times \nabla\times\left(
	\qvec + \frac{1}{\tau_1}\int_0^T\qvec\,\rmd T' \right)
	- \frac{3\ell_L^2}{\tau_1'} \nabla (\nabla\cdot\qvec)
	+ \frac{1}{\tau_1'} \qvec
	= 
	-\frac{1}{\tau_2'}\mathcal{R}(\Jmicrovec) ,
\end{equation}
which has to be compared with the original Guyer--Krumhansl equation
\cite{Guyer1966} on the heat flux $\qvec$ (e.g. see
\cite{EIT2010,Van2017,Zhu2017,dreyer1993}):
\begin{equation}\label{eq:GK.original}
	\pd_\Tsm \qvec + \frac{\kappa}{\tau'} \nabla\Ttot
	+ \frac{\ell^2}{\tau'} \big(
	 \nabla \times(\nabla\times\qvec)
	- 3 \nabla (\nabla\cdot\qvec)
	\big) 
	+ \frac{1}{\tau'} \qvec
	= 0.
\end{equation}

One can conclude that assuming a separation of time scales $\tau_1 \ll \tau_2$
(or equivalently $\Ch \gg \Ch$), the right hand side in \eqref{eq:GK.2T.q.int}
can be neglected, and the resulting equation is a generalization of the original
Guyer-Krumhansl equation \eqref{eq:GK.original} with two different length scales
$\ell_L$ and $\ell_T$ for longitudinal and transverse modes, respectively, and
an additional memory effect for the heat flux (the integral term in
\eqref{eq:GK.2T.q.int}). Moreover, the original GK value for the length scale
$\ell = \ell_L = \ell_T$ is obtained for $\sigma\Ch^2=3\eps\lambda^2/\mu_0$,
i.e. $\sigma/\eta=3/(\mu_0\Ch^2)$. Hence, for this to be true (i.e. $\ell_L =
\ell_T$), $\sigma$ and $\eta$ must be of the same order in $\lambda$. However,
in general their ratio is a free material constant (for $\sigma\gg\eps\lambda^2$
the longitudinal length dominates ($\ell_L\gg\ell_T$), and for
$\sigma\ll\eps\lambda^2$ the longitudinal term disappears).

Interestingly, that \eqref{eq:GK.2T.q.int} is structurally similar to the
generalized GK equation \cite[eq.(10)]{restuccia2026} obtained by Restuccia, Jou
and Pavelka \cite{restuccia2023,restuccia2026} within the EIT framework with two
extra tensorial state variables. In their generalized GK equation, there is also
an extra term which is likely linked to our $\Jtotcurl$-term. However, the two
approaches differe significantly in their foundations, and it would be
interesting to understand the precise relation between the two models in the
future

In summary, both nonlocal terms in the GK limit \eqref{eq:GK.2T}
come from the non-holonomy of the microscopic thermal impulse $\Jmicrovec$.
The transverse (curl-curl) term comes from its spatial curl through the thermal
vorticity $\BFvec=\alpha\nabla\times\Jmicrovec$. The longitudinal (grad-div)
term comes from its time component $A_\Osm=-\Tmicro$, i.e. from the non-holonomy
in time.

\section{Dispersion analysis of the two-temperature model}\label{sec:dispersion}

Denoting $\bm{w} = \{\Stot,\Jtotvec,\Smicro,\Jmicrovec,\DFvec,\BFvec\}$, we look
for plane wave solutions ($x = X^1$)
\begin{equation}\label{eq:wave}
	\bm{w} = \hat{\bm{w}} e^{i(k x - \omega t)},
\end{equation}
of the linearized equations \eqref{eq:3D.2temp.diss}
\begin{equation}
   \pd_\Tsm \bm{w} + \mathbb{C} \pd_x \bm{w} = \mathbb{S}\bm{w}
\end{equation}
near the equilibrium state state $\Ttot_0 = 1$, $S=S_0$, $s=0$,
$\Jtotvec=\Jmicrovec = 0$, $\BFvec = 0$, $\DFvec = 0$. Here, $\hat{\bm{w}}$ is
the constant amplitude vector, $k$ is the wave number, $\omega$ is the angular
frequency, and $i = \sqrt{-1}$. After substituting the ansatz \eqref{eq:wave}
into the linearized equations. We obtain an eigenvalue problem:
\begin{equation}
	\left( k \mathbb{C} + i \mathbb{S} - \omega \mathbb{I} \right) \hat{\bm{w}} = 0,
\end{equation}
or
\begin{equation}
	\left( \mathbb{C} + \frac{i}{k} \mathbb{S} - \nu \mathbb{I} \right) \hat{\bm{w}} = 0,
\end{equation}
where $\nu = \omega/k$, and $\mathbb{C}$ and $\mathbb{S}$ are Jacobians of the
fluxes and source terms, respectively, evaluated at the equilibrium state, e.g.
see \cite{peshkov2026a}, while $\mathbb{I}$ is the identity matrix.

This eigenvalue problem has nontrivial solutions if and only if
\begin{equation}\label{eqn.dispersion.relation}
	p(k,\nu) := \det \left( \mathbb{C} + \frac{i}{k} \mathbb{S} 
	- \nu \mathbb{I} \right) = 0.
\end{equation}
Then, for a given real frequency $\omega$ and complex wave number $k$, the phase
velocity $\Vph(\omega)$ and the attenuation factor $a(\omega)$ are defined as
\begin{equation}
	\Vph(\omega) = \frac{\omega}{\mathrm{Re}(k)}, 
	\qquad
	a(\omega) = \mathrm{Im}(k),
\end{equation}
while for a given real wave number $k$ and complex frequency $\omega$, the phase
velocity $\Vph(k)$ and the damping rate $\gamma(k)$ are defined as
\begin{equation}
	\Vph(k) = \mathrm{Re}(\nu), 
	\qquad
	\gamma(k) = \mathrm{Im}(\nu).
\end{equation}

We shall distinguish three different modes that correspond to the three
different types of eigenvalues of the hyperbolic part of the model, i.e. the
eigenvalues of the flux Jacobian $\mathbb{C}$:
\begin{subequations}\label{eq:modes.hyperb}
	\begin{equation}\label{eq:mode.vort}
		C_{\mathrm{vort}}^2 = \frac{1}{\eps_0 \mu_0},
	\end{equation}
	\begin{equation}\label{eq:mode.fast}
		C_{\mathrm{fast}}^2 = \frac{\Ch^2 (c_s+c_S)+c_S \ch^2 + 
		\sqrt{\Ch^4 (c_s+c_S)^2+2 c_S \Ch^2 \ch^2 (c_S-c_s)+c_S^2 \ch^4}}{2c_s c_S},	
	\end{equation}
	\begin{equation}\label{eq:mode.slow}
		C_{\mathrm{slow}}^2 = \frac{\Ch^2 (c_s+c_S)+c_S \ch^2 - 
		\sqrt{\Ch^4 (c_s+c_S)^2+2 c_S \Ch^2 \ch^2 (c_S-c_s)+c_S^2 \ch^4}}{2c_s c_S}.
	\end{equation}	
\end{subequations}
We shall call \eqref{eq:mode.vort} the \emph{thermal vorticity} mode,
\eqref{eq:mode.fast} the \emph{fast thermal} mode (Cattaneo mode), and
\eqref{eq:mode.slow} the \emph{slow thermal} mode. 

\begin{figure}[htbp]
  \centering
  \begin{minipage}[b]{0.28\textwidth}
	\centering
	\includegraphics[width=\linewidth]{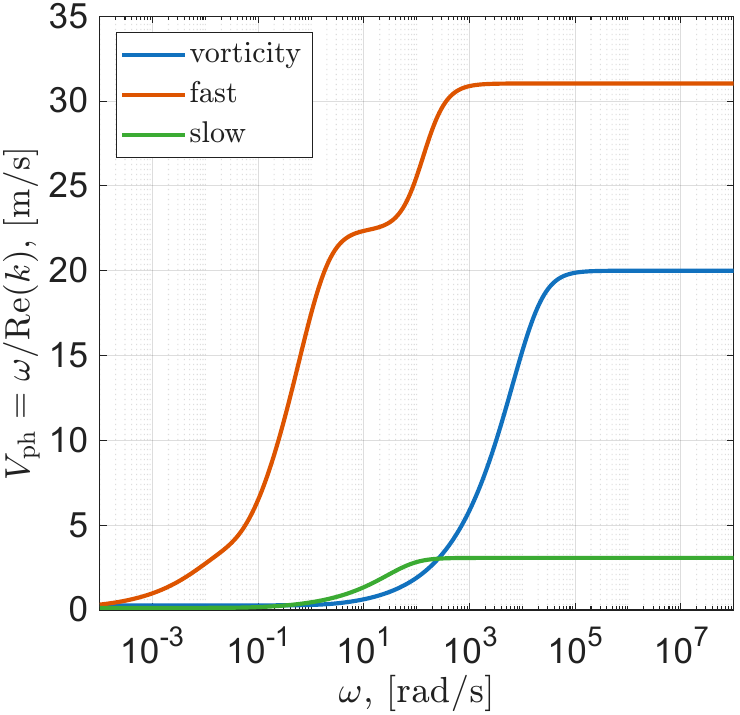}
	{\small (a) Phase velocities $\Vph(\omega)$.}
  \end{minipage}
  \hspace{0.5cm}
  \begin{minipage}[b]{0.28\textwidth}
    \centering
    \includegraphics[width=\linewidth]{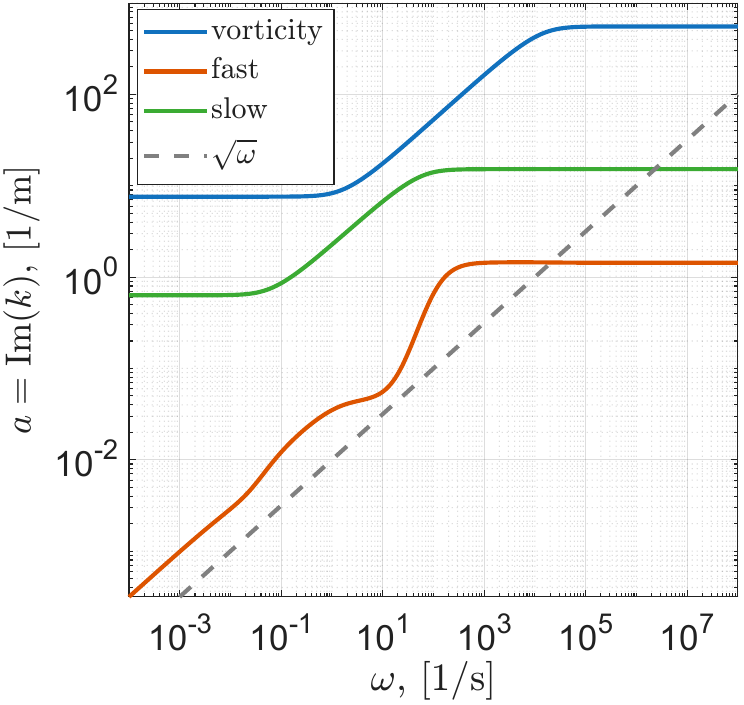}
	{\small (b) Attenuation factor $a(\omega)$.}
\end{minipage}
\hspace{0.5cm}
\begin{minipage}[b]{0.28\textwidth}
    \centering
    \includegraphics[width=\linewidth]{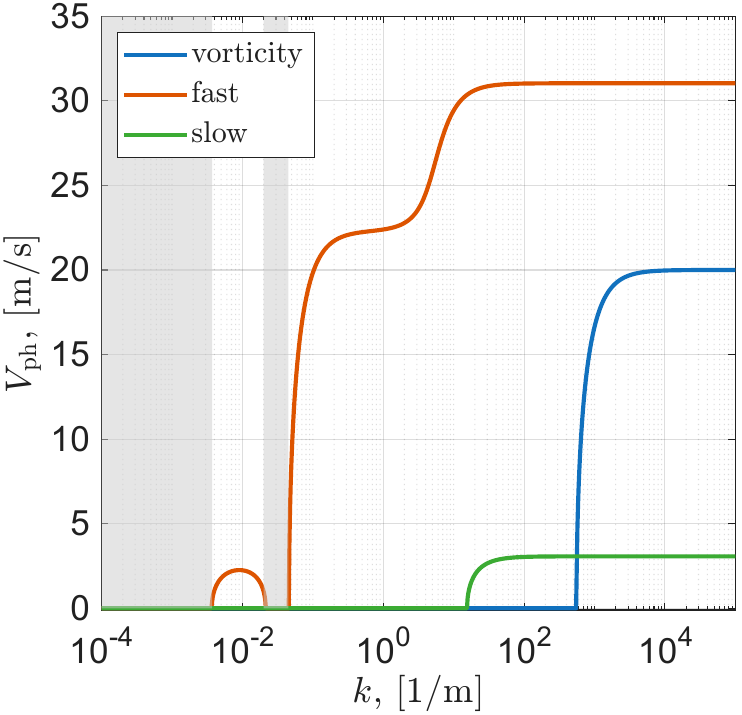}
	{\small (c) Phase velocities $\Vph(k)$.}
  \end{minipage} 
  \caption{Typical dispersion curves for the two-temperature model when all
  relaxation terms are active. The material parameters are given in
  Table~\ref{tab:dispersion}. The gray zones in (c) correspond to the $k$-band
  gaps.}
  \label{fig:dispersion}
\end{figure}

\begin{figure}[htbp]
  \centering
  \begin{minipage}[b]{0.28\textwidth}
	\centering
	\includegraphics[width=\linewidth]{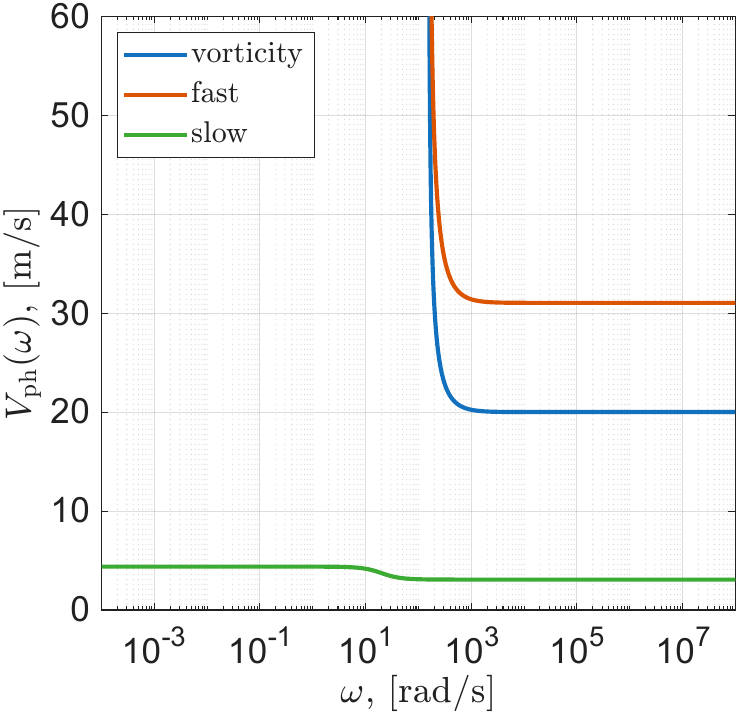}
	{\small (a) Phase velocities $\Vph(\omega)$.}
  \end{minipage}
  \hspace{0.5cm}
  \begin{minipage}[b]{0.28\textwidth}
    \centering
    \includegraphics[width=\linewidth]{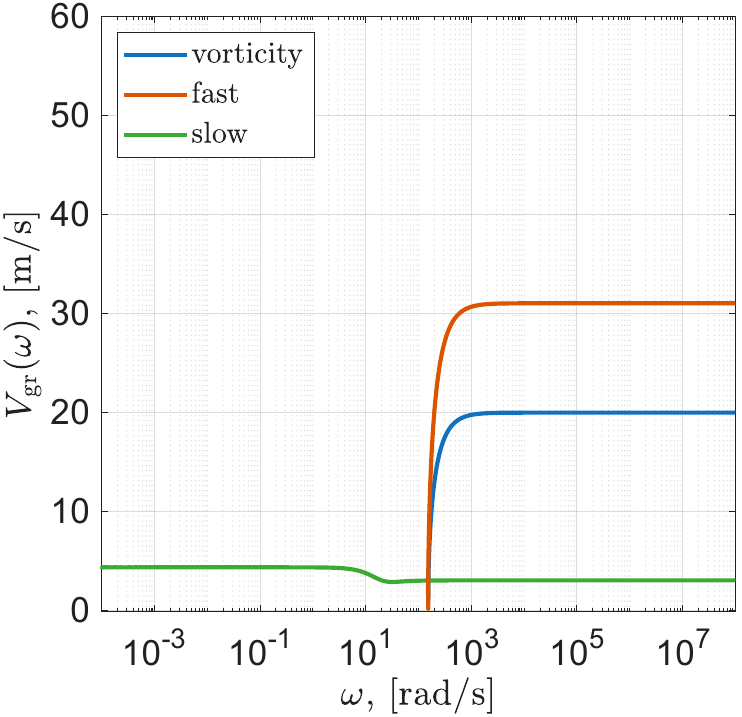}
	{\small (b) Group velocities $\Vgr(\omega)$.}
\end{minipage}
  \caption{Typical dispersion curves for the two-temperature model when all the
  dissipative relaxation terms are switched off (ballistic regime), i.e. $\eps =
  \eta = \sigma = \infty$. Other material parameters are given in
  Table~\ref{tab:dispersion}.}
  \label{fig:dispersion.ballistic}
\end{figure}

In Fig.\,\ref{fig:dispersion}, we show the typical dispersion curves for the
two-temperature model when all relaxation terms are active. The material
parameters used for the dispersion curves are given in
Table~\ref{tab:dispersion}. We note that the phase velocities $\Vph(\omega)$ and
$\Vph(k)$ converge to the characteristic speeds of the hyperbolic part of the
model \eqref{eq:modes.hyperb} when $\omega \to \infty $ (or $k \to \infty $),
e.g. see \cite{Ruggeri1992}. Also note that in Fig.\,\ref{fig:dispersion}(b),
the attenuation factor $a(\omega)$ of the fast thermal mode (Cattaneo mode)
relaxes at the rate $\varpropto \sqrt{\omega}$ that corresponds to the classical
diffusion (Fourier) limit of the Cattaneo heat conduction.
Fig.\,\ref{fig:dispersion}(c) shows that there are two $k$-band gaps in the
phase velocity $\Vph(k)$, while no frequency band gap is present in
$\Vph(\omega)$ in Fig.\,\ref{fig:dispersion}(a). This is not a contradiction,
because the two representations correspond to different physical problems. For
real $k$ and complex $\omega$ (initial-value problem), a $k$-gap is an interval
of wave numbers in which $\mathrm{Re}(\omega)=0$. The corresponding
perturbations do not propagate but decay monotonically in time, i.e. they are
overdamped. Such gapped momentum states are typical of hyperbolic relaxation
models, e.g. \cite{baggioli2020}. For real $\omega$ and complex $k$
(boundary-value problem), the dissipative source terms give
$\mathrm{Re}(k)\neq0$ for all $\omega>0$. Dissipation therefore does not produce
a frequency band gap in the strict sense, i.e. an interval of $\omega$ with
$\mathrm{Re}(k)=0$ (evanescent waves). It only produces strong attenuation, see
Fig.\,\ref{fig:dispersion}(b). Genuine frequency gaps are instead associated
with the reversible part of the model. Thus, in the ballistic regime of
Fig.\,\ref{fig:dispersion.ballistic} (all dissipative relaxation terms are
turned off $\eps = \eta = \sigma = \infty$), the thermal vorticity mode and the
fast thermal mode have cut-off frequencies, at which $\Vph(\omega)\to\infty$ and
$\Vgr(\omega)\to0$. Below them, $k$ is purely imaginary and these modes are
evanescent. This is the same mechanism that produces frequency band gaps in
microstructured solids \cite{Madeo2015a,peshkov2026a}. Here, the cut-offs are
produced by the antisymmetric $\alpha^{-1}$-coupling, which acts as a mass term
for the thermal vorticity. When the dissipation is switched on, the sharp
cut-offs are smeared into regions of strong attenuation, while $k$-gaps appear.

\begin{table}[htbp]
\centering
\small
\renewcommand{\arraystretch}{1.}
\begin{tabularx}{\textwidth}{>{\raggedright\arraybackslash}p{3.2cm} >{\raggedright\arraybackslash}X >{\centering\arraybackslash}p{3.0cm}}
	\toprule
	Symbol & Meaning & Value\\
	\midrule
	$\lambda$ & length-scale parameter & $3\times10^{-3}$\\
	$\sigma$ & relaxation parameter of $\Smicro$ & $10^{-5}$\\
	$\Ch$ & macroscopic thermal-wave speed parameter & $500$\\
	$\ch$ & microscopic thermal-wave speed parameter & $100$\\
	$\tau_1$ & relaxation time scale of $\Jtotvec$ & $1$\\
	$C_\mathrm{vort} = \frac{1}{\sqrt{\eps_0\mu_0}}$ & thermal vorticity speed & $20$\\
	$\eps_0$ & thermal permittivity parameter & $2\times10^{-5}$\\
	$c_S$ & specific heat of the observed entropy & $500$\\
	$c_s$ & specific heat of the fluctuating entropy & $550$\\
	\midrule
	\multicolumn{3}{l}{{Derived quantities}}\\
	$\eps = \Ch^2\tau_1$ & relaxation parameter of $\Jtotvec$ & $2.5\times10^{5}$\\
	$\alpha = \lambda\eps$ & length-scale parameter & $750$\\
	$\eta = \lambda^2\eps$ & thermal resistivity coefficient & $2.25$\\
	$\mu_0 = 1/(\eps_0 C_\mathrm{vort}^2)$ & thermal permeability parameter & $125$\\
	$C_\mathrm{fast}$ & fast mode characteristic speed & $31.037$\\
	$C_\mathrm{slow}$ & slow mode characteristic speed & $3.072$\\
	\bottomrule
\end{tabularx}
\caption{Material parameters used for the dispersion curves in
  Figs.~\ref{fig:dispersion} and \ref{fig:dispersion.ballistic}. In the
  ballistic case of Fig.~\ref{fig:dispersion.ballistic}, $\eps=\eta=\sigma=\infty$.}
\label{tab:dispersion}
\end{table}

In Fig.\,\ref{fig:dispersion.ballistic}, we show the typical dispersion curves
for the two-temperature model when all the dissipative relaxation terms are
switched off (ballistic regime), i.e. $\eps = \eta = \sigma = \infty$. The
remaining active relaxation terms are the antisymmetric source terms with the
$\alpha^{-1}$ factor. One can see in Fig.\,\ref{fig:dispersion.ballistic}(a)
that the phase velocities $\Vph(\omega)$ corresponding to the thermal vorticity
mode and the fast thermal mode are unbounded. However, there is no contradiction
with the causality principle because the energy is transmitted with the group
velocities $\Vgr(\omega)$, which are bounded, as can be seen in
Fig.\,\ref{fig:dispersion.ballistic}(b).

\section{Summary and discussion of the two-temperature
model}\label{sec:discussion}

Macroscopic modeling of non-equilibrium multiscale phenomena remains a
challenging task. Many existing models rely on a considerable number of
phenomenological assumptions. This makes them difficult to generalize and to
couple consistently with other physical processes. Over the past decades,
several frameworks based on general principles have been developed to address
this problem, notably GENERIC \cite{GrmelaOttingerI,GrmelaOttingerII} and
Rational Extended Thermodynamics \cite{MullerRuggeri1998,Ruggeri2021a}. These
frameworks constrain mainly the \emph{dynamics} of the non-equilibrium
variables, through Hamiltonian structure, entropy production and hyperbolicity.
Much less attention has been paid to the \emph{kinematics} of these variables,
i.e. to the geometric nature of the fields that describe the unresolved degrees
of freedom. The differential geometric structure that underlies these fields is,
however, what separates the universal part of a theory from its
material-specific part. The kinematic identities that follow from $\rmd^2=0$
hold independently of the constitutive assumptions. The material information, in
contrast, enters only through the Hamiltonian and the relaxation terms. The
geometric structure also determines which couplings between fields are
admissible, and which constraints have to be preserved by a numerical scheme. In
this sense, geometry provides a physics-independent and internally consistent
framework that complements the existing thermodynamic principles. In this paper,
we have shown how this idea can be realized for heat conduction, by constructing
a formulation of its non-equilibrium thermodynamics and dynamics in terms of
differential forms on the 4D matter-time manifold. In this section, we summarize
the main results of the paper and discuss the physical significance of the
geometric framework, as well as its implications for multiphysics modeling and
for structure-preserving numerical methods.

\subsection{Hierarchy of models}\label{ssec:hierarchy}

We have developed a non-equilibrium heat conduction model that has a
hierarchical structure, i.e. it contains the simpler models of heat conduction,
the Fourier and Cattaneo, as relaxation limits. Each level is obtained from the
next one by the relaxation of the additional fields, i.e. by eliminating the
fast degrees of freedom \cite{Grmela2015,PKG_Book2018,Crespo2025a}. Moving up
the hierarchy, each level adds a geometric structure: the base model adds the
variational (Hamiltonian) structure on the Fourier level of description, and the
two-temperature model adds the exterior-calculus structure
($\F{}=\rmd\Jmicrovec$, $\rmd\F{}=0$) to the base model. The hierarchy is
summarized in Table~\ref{tab:hierarchy}.
\begin{table}[h!]
\centering
\small
\renewcommand{\arraystretch}{1.3}
\begin{tabularx}{\textwidth}{>{\raggedright}p{3.0cm} >{\raggedright}p{2.3cm} >{\raggedright}X >{\raggedright\arraybackslash}p{3.4cm}}
	\toprule
	Model & State variables & Geometric structure & Relaxation limit\\
	\midrule
	Fourier & $\Stot$ ($\Ttot$) & none; parabolic, $\Jtotvec\approx-\tau\nabla\Ttot$ is not an independent field & ---\\
	Base model (Cattaneo type), Sec.~\ref{sec:base} & $\Stot$, $\Jtotvec$ &
	variational; $\Jtotvec=\rmd\varphi$ (reversible); $\F{}=\rmd\Jtotvec\neq0$ is produced by dissipation, but is not an independent field & $\tau\to0$: Fourier\\
	Two-temperature model, Sec.~\ref{sec:ext.model} &
	$\Stot$, $\Jtotvec$, $\Smicro$, $\Jmicrovec$, \\ $\DFvec$, $\BFvec$ &
	variational + exterior calculus; $\Jmicrovec$ non-holonomic, $\F{}=\rmd\Jmicrovec$ independent with $\rmd\F{}=0$; Maxwell-type structure &
	$\lambda\to0$: GK-type, Sec.~\ref{sec:GK2}; then $\tau\to0$: Fourier\\
	\bottomrule
\end{tabularx}
\caption{Hierarchy of heat-conduction models contained in the two-temperature
model \eqref{eq:3D.2temp.diss}.}
\label{tab:hierarchy}
\end{table}

By degrees of freedom comparison, our model is similar to the one by Dreyer and
Struchtrup \cite{dreyer1993}. In their classification, the hierarchy of heat
conduction regimes can be understood as the presence and interaction of
three types of regimes: diffusive phonons, second sound, and ballistic phonons.
In our model, in terms of the fields $\Jtotvec$, $\Jmicrovec$ and $\F{}$, this
hierarchy of regimes can be summarized as follows:
\begin{itemize}
	\item 
	In the Fourier
	model, $\Jtotvec$, $\Jmicrovec$ and $\F{}$ are not
	independent fields: only \emph{diffusive phonons}. 
	\item
	In the base model, $\Jtotvec$ is an independent field, while $\Jmicrovec=0$:
	\emph{diffusive phonons} and \emph{second sound}. 
	\item
	In the two-temperature model, all fields $\Jtotvec$, $\Jmicrovec$ and $\F{}$
	are independent: \emph{diffusive phonons}, \emph{second sound}, and
	\emph{ballistic phonons}.
\end{itemize}

\subsection{Physical context}\label{ssec:physical.context}

Let us now summarize the physical significance of the new fields in the
two-temperature model, and the physical situations in which they are expected to
be relevant.

The two new vector fields $\BFvec$ and $\DFvec$ have a direct kinematic
interpretation. The magnetic-like field $\BFvec=\alpha\nabla\times\Jmicrovec$ is
the spatial \emph{thermal vorticity}: the circulation of the small-scale thermal
impulse per unit area. It is non-zero when the unresolved heat carriers move
along closed, vortex-like paths, which leave no trace in the total impulse
$\Jtotvec$. Its energy $\Vert\BFvec\Vert^2/2\mu_0$ is the energy stored in such
circulating micro-flows. In the Guyer--Krumhansl limit, it produces the
transverse (curl--curl) term.

The electric-displacement-like field $\DFvec$ is the \emph{thermal displacement
field}, its role is to describe the microinertia of unresolved scales
\cite{peshkov2026a} via its dual $\EFvec = -\alpha\left(\pd_\Tsm\Jmicrovec +
\nabla\Tmicro\right)$ (the non-holonomy of $\Jmicrovec$ in time, see
Section~\ref{ssec:geom.setup}). As we have seen, from the Gauss-type law
$\nabla\cdot\DFvec=-\alpha^{-1}\Smicro$ in the reversible case, the fluctuating
entropy is its source.

Moreover, the fields $\BFvec$ and $\DFvec$ are related to the \emph{thermal
Poynting vector} $\EFvec\times\HFvec$ in \eqref{eq:q.ext} via their duals
$\EFvec$ and $\HFvec$. The thermal Poynting vector is a contribution to the heat
flux that is carried neither by the observable nor by the small-scale impulse,
but by the interplay of the two components of the thermal vorticity. It is
quadratic in the non-equilibrium fields, so it does not appear in the linear GK
limit. It becomes relevant for strongly non-equilibrium, nonlinear heat
transport, e.g. for heat flux circulating around obstacles, where it can
transport energy perpendicular to both $\EFvec$ and $\HFvec$
\cite{Sykora2023,ziabari2018,zhang2021}.

Overall, non-equilibrium and non-local effects described by $\DFvec$ and
$\BFvec$ are expected to be important whenever the mean free path of the heat
carriers is comparable with the geometric length scales, or the relaxation times
are comparable with the time scales of the process. Thus, we expect potential
applications might be in the following areas.

\emph{1) Nanomaterials and nanodevices.} In thin films, nanowires and
nanolayers, the phonon mean free path can exceed the sample size. Heat
transport is then non-local, with boundary slip of the heat flux and
Knudsen-type size effects, see \cite{Zhu2017,zhang2021}.
GK-type models have been shown to be useful as a continuum description in this
regime \cite{Van2017,EIT2010,Sykora2023}.

\emph{2) Phonon hydrodynamics and thermal vortices.} When normal
(momentum-conserving) phonon collisions dominate resistive ones, phonons behave
as a viscous fluid \cite{Guyer1966,Sykora2023,zhang2021,restuccia2026}, and
vortical heat flux patterns become possible. Thermal vortices have recently been
studied within a multiscale GENERIC framework \cite{Sykora2023}. In the present
model, such vortices are described by $\BFvec$, and their evolution is governed
by the kinematic identity $\rmd\F{}=0$. Recently, Di Lucente et
al.~\cite{dilucente2026} showed that the steady viscous heat equations of phonon
hydrodynamics, which are obtained by coarse-graining the linearized phonon
Boltzmann equation, decouple into separate (modified Helmholtz) equations for
the compressibility $\nabla\cdot\bm{u}$ and the vorticity $\nabla\times\bm{u}$
of the phonon drift velocity $\bm{u}$. The temperature field is then the sum of
a compressible and a vortical contribution. In a graphite strip, the vortical
part produces thermal vortices and heat backflow (a negative non-local thermal
resistance), and this effect is strongest for nearly incompressible phonon flow.
A similar separation is built into the present model. The longitudinal
(grad--div) mode is controlled by the fluctuating temperature $\Tmicro$, i.e.
by the non-holonomy of $\Jmicrovec$ in time, and the transverse (curl--curl)
mode is controlled by the thermal vorticity
$\BFvec=\alpha\nabla\times\Jmicrovec$, i.e. by its non-holonomy in space. The
two channels have independent length scales $\ell_L$ and $\ell_T$, see
Section~\ref{sec:GK2}. The strip geometry of \cite{dilucente2026} is therefore a
natural benchmark for the present model. Such a comparison, however, requires
the full system \eqref{eq:3D.2temp.diss} with suitable boundary conditions for
$\Jmicrovec$, $\DFvec$ and $\BFvec$, rather than its GK limit, because in the
latter the transverse term contains the memory integral in
\eqref{eq:GK.2T.q.int}.
	
\emph{3) Superfluids.} In superfluid helium, heat is transported by the
counterflow of the normal and superfluid components \cite{khalatnikov2018}, and
quantized vortices interact with the heat flux \cite{mongiovi2018}. The
two-temperature structure (an observable and a fluctuating temperature) and the
thermal vorticity are natural ingredients for such systems.
	
\emph{4) Microstructured materials and thermal metamaterials.} In materials with
an internal length scale (lattices, foams, composites), the micro-scale heat
flux can differ from the effective macroscopic one. This is in analogy with
micromorphic continua in mechanics \cite{neff2014a,Madeo2015a,peshkov2026a},
where the micro-distortion is not compatible with the macroscopic displacement.
Room-temperature non-Fourier behaviour, e.g. GK-type or other non-local effects,
have been observed in heterogeneous materials
\cite{Van2017,kovacs2024,ganghoffer2025}.

\emph{5) Turbulent flows.} A turbulent flow can be viewed as a medium with
emergent microstructure, in which the unresolved eddies carry part of the heat.
The fluctuating temperature $\Tmicro$ and the small-scale impulse $\Jmicrovec$
can then be interpreted as subgrid (unresolved) temperature fluctuations and
turbulent heat flux, and $\BFvec$ as their vorticity. Coupling of the present
model with the geometric unified model of continuum mechanics \cite{PRD2019}
might be an interesting direction for future research in turbulence modeling.

\subsection{Is this a gauge theory?}\label{ssec:gauge}

The proposed theory has the \emph{kinematics} \eqref{eq:F.oneform} of an abelian
gauge theory but not its \emph{gauge symmetry} as we have already mentioned in
Section~\ref{rem:whyF}. A gauge theory has two key aspects: a bundle geometry,
and a redundancy of description expressed as invariance under local
transformations of the fiber. In our model, we adopt the first and deliberately
reject the second. This is done for a physical reason rather than a technical
one. Indeed, we want the proper temperature $\hat{\Theta}=\sqrt{-\Jtot_\Asm
J^\Asm }$ to be a measurable field, and a local shift symmetry in the thermal
impulse $\Jtot_\Asm\to \Jtot_\Asm+\pd_\Asm\chi(\XX)$ would declare it
unobservable. Therefore, strictly speaking, the two-temperature model is not a
gauge theory. However, as we discuss in what follows, it can be seen as a gauge
theory with a broken symmetry, and the symmetry is effectively restored in the
near-equilibrium limit.

In the geometric setup of the reversible two-temperature model,
Section\,\ref{ssec:geom.setup}, the decomposition \eqref{eq:J.oneform} is
invariant under the local transformation of the thermal displacement potential
$\varphi$:
\begin{equation}\label{eq:J.gauge}
	\varphi \to \varphi + \chi(X^\Bsm),
	\qquad
	\Jmicro_\Bsm \to \Jmicro_\Bsm + \pd_\Bsm\chi,
	\qquad
	\Jlarge_\Bsm = \Jtot _\Bsm - \Jmicro_\Bsm \to \Jlarge_\Bsm,
\end{equation}
i.e. $\Jlargevec$ and $-\rmd\Jlargevec = \rmd \Jmicrovec = \F{}$ do not change.
Formally, if the Lagrangian depends on $\Jlargevec$ and $\F{}$ only, then
\eqref{eq:J.gauge} is a local symmetry (gauge symmetry). In this case, the
reversible part of the model has the geometry of an abelian gauge theory,
$\varphi$ is a section of a bundle with structure group $(\mathbb{R},+)$,
$\Jmicrovec$ is a connection on it, $\Jlargevec = \rmd\varphi - \Jmicrovec$ is
the covariant derivative of $\varphi$, and $\F{} = \rmd\Jmicrovec$ is the
curvature. Moreover, by Noether's second theorem the local symmetry implies an
identity between the field equations (Bianchi-type): in our case, it is the
$\varphi$-equation \eqref{eq:4D.ext.s} $\pd_\Bsm L_{\Jmicro_\Bsm} = 0$ that is
the divergence of the $\Jmicrovec$-equation.

However, under \eqref{eq:J.gauge}, the two temperatures transform as
\begin{equation}\label{eq:T.gauge}
	\Ttot \to \Ttot - \pd_\Tsm\chi,
	\qquad
	\Tmicro \to \Tmicro - \pd_\Tsm\chi,
	\qquad
	\Tlarge = \Ttot - \Tmicro\to\Tlarge .
\end{equation}
A gauge-invariant Lagrangian can therefore depend on the time components only
through $\Tlarge=-\Jlarge_\Osm$. This results in
$\Stot=L_\Ttot=L_\Tlarge=L_\Tmicro=\Smicro$, i.e. the two entropies coincide,
and the two entropy equations \eqref{eq:3D.ext.Sobs} and
\eqref{eq:3D.ext.Smicro} become identical. The model with such a Lagrangian,
however, converges to a GK-type equation without the longitudinal mode, and
therefore was rejected. Conversely, the two-temperature model requires a
Lagrangian that depends on $\Ttot$ and $\Tmicro$ separately. For example, the
Hamiltonian \eqref{eq:H.quadratic.GK2} has separate energies $\veps(\Stot)$ and
$\Smicro^2/2c_\Smicro$. This, in turn, breaks the local symmetry
\eqref{eq:J.gauge} explicitly. Similarly, the spatial part of
\eqref{eq:H.quadratic.GK2} depends not only on invariant fields $\Jlargevec$ and
$\BFvec$, but it has explicit dependence on $\Jmicrovec$ (the term $\ch^2/2
\Vert\Jmicrovec\Vert^2$), which also breaks the local symmetry
(similar to massive field theories of Proca-type).

Therefore, the two-temperature model can be seen as a gauge theory in which the
symmetry \eqref{eq:J.gauge} is broken. The symmetry is formally restored only in
the near-equilibrium limit, when  $\Tmicro\to0$ and $\Jmicrovec\to0$ and the
symmetry breaking terms become inactive.

Finally, let us comment about the similarity of the two-temperature model to a
Maxwell-type field theory. The Maxwell-type form of \eqref{eq:3D.ext} is not a
coincidence or an analogy imposed by hand. In fact, any first-order field theory
of a $1$-form potential on a four-dimensional manifold, with a Lagrangian
depending on the potential and on its exterior derivative $\F{}$, leads to the
same structure \cite{Olivares2022,peshkov2026a,Peshkov2025}. This is because the
kinematic identity $\rmd\F{} = 0$ gives the homogeneous pair
($\BFvec$-equation and $\nabla\cdot\BFvec=0$), independently of the dynamics
(constitutive relations). Then, the structure of the Euler-Lagrange equation
$\pd_\Asm \Hexci{\Bsm\Asm} = -L_{\Jmicro_{\Bsm}}$, as we have seen, is also
rather rigid and gives the inhomogeneous pair (the $\DFvec$-equation and the
Gauss-type law) as in other fields \cite{Olivares2022,peshkov2026a,Peshkov2025}.
And finally, in four dimensions, the Hodge dual of a $2$-form is again a
$2$-form. This makes the $3+1$ split of $\F{}$ and $\FStar{}$ produce two
spatial vector fields each, $(\EFvec,\BFvec)$ and $(\DFvec,\HFvec)$.

The material information enters only through the constitutive relations
$\EFvec=\Ham_{\DFvec}$, $\HFvec=\Ham_{\BFvec}$, i.e. through the Hamiltonian
$\Ham$. This is the same universality that makes the equations of
electrodynamics, of the teleparallel formulation of gravity
\cite{Peshkov2025,Olivares2022}, of the gauge theory of dislocations
\cite{KadicEdelen1983,Lazar2000}, and of continuum mechanics with torsion
\cite{PRD2019,Hehl2007} look alike. In all these cases the governing equations
are the field equations of a connection (potential) whose curvature or torsion
is the field strength. The specific physics of heat conduction is in the
Hamiltonian and in the coupling term $\frac{1}{2}\Ch^2\Jlargevec^2$, which makes
the theory massive (Proca-like) rather than Maxwell-like.

\subsection{Bundle geometric framework}\label{ssec:bundles}

\subsubsection{Hierarchy of bundles}
The construction of Section~\ref{sec:ext.model} is an instance of a general
procedure. A bundle is itself a smooth manifold, and another bundle can be
attached to it. The base model uses the trivial line bundle
$\mathscr{M}\times\mathbb{R}$, whose sections are the thermal displacement
$\varphi$, and its first jet bundle, whose fiber coordinates are $\Jtot_\Asm$. The
two-temperature model adds a connection $\Jmicrovec$ on this bundle, with
curvature $\F{}$.

A naive way to continue would be to raise the form degree, i.e. to promote the
non-holonomy of $\F{}$ to an independent $2$-form potential with a $3$-form
field strength. On the four-dimensional manifold $\mathscr{M}$, however, a
hierarchy of scalar-valued forms is necessarily finite. A $3$-form potential has
a top-degree ($4$-form) field strength, which carries no local dynamics in the
massless case, and a $4$-form potential has a vanishing field strength.
Moreover, by Hodge duality in four dimensions, $2$- and $3$-form fields are
equivalent to the vector and scalar fields already present. The hierarchy can
nevertheless be continued in two other directions, in which the bound on the
form degree does not apply.

\paragraph{Forms with values in larger fibers.}
Instead of the form degree, one can increase the number of components of the
fiber in which the forms take values. The scalar-valued $1$-form $\Jmicro_\Bsm$
is then replaced by a vector- or tensor-valued $1$-form $\Jmicro^a_{\ \Bsm}$,
$a=1,\dots,N$, i.e. by a connection on a bundle with an $N$-dimensional fiber
(or with an $N$-dimensional structure group). Its curvature
\begin{equation*}
	\F{\ \Asm\Bsm}^{a} = \pd_\Asm \Jmicro^a_{\ \Bsm} - \pd_\Bsm \Jmicro^a_{\ \Asm}
	+ c^a_{\ bc}\,\Jmicro^b_{\ \Asm} \Jmicro^c_{\ \Bsm},
	\qquad
	\text{or}
	\qquad
	\F{}^a = \rmd\Jmicrovec^a + \tfrac{1}{2}c^a_{\ bc}\,\Jmicrovec^b\wedge\Jmicrovec^c,
\end{equation*}
contains the structure constants $c^a_{\ bc}$ of the structure group if this
group is non-abelian. It satisfies the Bianchi identity $\mathrm{D}\F{}^a=0$,
with $\mathrm{D}$ the covariant exterior derivative, which reduces to
$\rmd\F{}^a=0$ in the abelian case. The $3+1$ split then gives $N$ copies of the
Maxwell-type pairs $(\EFvec^a,\BFvec^a)$ and $(\DFvec^a,\HFvec^a)$, which are
coupled through the quadratic term in the non-abelian case. This is the route
taken in microstructured continua \cite{peshkov2026a,Crespo2025a}, where the
micro-distortion is a tensor-valued $1$-form (also called solder form), and in
the two-scale fiber-bundle framework of \cite{Crespo2025a}. In heat conduction,
the fiber index $a$ can label, for example, different groups of heat carriers
(phonon branches or polarizations), each with its own fluctuating temperature
$\Tmicro^a=-\Jmicro^a_{\ \Osm}$ and its own relaxation time.

\paragraph{Forms on the total space of a bundle and on jet bundles.}
The second route uses the fact that a bundle is itself a manifold. Let
$\pi:\mathscr{E}_1\to\mathscr{M}$ be a fiber bundle with a $k_1$-dimensional
fiber, e.g. the bundle whose fiber coordinates are the fields of the current
level of the hierarchy. Its total space has dimension $4+k_1$. Differential
forms on $\mathscr{E}_1$ can therefore have degree up to $4+k_1$, and a new
connection, its curvature and its Bianchi identity can be defined on
$\mathscr{E}_1$ itself. The splitting of forms on $\mathscr{E}_1$ into
horizontal components (along $\mathscr{M}$) and vertical components (along the
fiber) gives fields that depend not only on the matter-time point but also on
the fiber coordinates, i.e. on the state of the previous level. This is
analogous to kinetic theory, where the distribution function lives on phase
space rather than on space-time. The same effect is obtained by passing to
higher-order jet bundles $J^r\mathscr{E}$, whose dimension grows with the order
$r$. Repeating the construction, i.e. attaching a bundle to the total space of
the previous one, produces a hierarchy
\begin{equation*}
	\mathscr{M}
	\;\longleftarrow\;
	\mathscr{E}_1
	\;\longleftarrow\;
	\mathscr{E}_2
	\;\longleftarrow\;\cdots,
	\qquad
	\dim\mathscr{M} < \dim\mathscr{E}_1 < \dim\mathscr{E}_2 < \cdots,
\end{equation*}
in which the admissible form degree increases at each step.

Along either route, each new level describes finer-scale non-equilibrium degrees
of freedom and comes with its own Bianchi identity, i.e. with its own pair of
dynamical and kinematic equations in the SHTC form. Each level relaxes to the
previous one in the corresponding relaxation limit, in the same way as the
two-temperature model relaxes to the base model (Section~\ref{ssec:hierarchy}).
This is similar to the hierarchy of models in extended non-equilibrium
thermodynamics theories \cite{Grmela2015,PKG_Book2018,MullerRuggeri1998}.

\subsubsection{Coupling of fields in multiphysics problems}

In multiphysics applications, one of the key questions is how different physical
fields couple to each other in the specification of the Lagrangian. The
geometric framework restricts the possible couplings to a small number of
universal types. 
\begin{itemize}
	\item \emph{Minimal coupling}: a potential enters only through a
	covariant combination, e.g. as $\Jlargevec = \rmd\varphi - \Jmicrovec$.
	\item \emph{Coupling through the curvature}: the Lagrangian depends on
	$\F{}$, as in Section~\ref{sec:ext.variational}.
	\item \emph{Coupling through the geometry of $\mathscr{M}$}: the Hodge star,
	and hence the constitutive relations $\DFvec(\EFvec)$, $\HFvec(\BFvec)$,
	depend on the metric. The thermal vorticity therefore couples to the strain
	of the material manifold. If the matter sector has torsion, $\F{}$ acquires
	the term $\Tors{\Csm}{\Asm}{\Bsm}\Jmicro_\Csm$ (Section~\ref{rem:torsion}),
	which couples heat conduction to defects, or gravity
	\cite{Peshkov2025,Olivares2022,AldrovandiPereiraBook} already at the
	kinematic level.
	\item \emph{Mixed and topological couplings}: terms such as
	$\F{}\wedge\F{}'$ between the thermal vorticity and another $2$-form field
	strength $\F{}'$ (e.g. the electromagnetic field, gravity via torsion). In
	the abelian case, the pure term $\F{}\wedge\F{}$ is a total derivative and
	does not change the field equations. Mixed terms between different fields,
	however, provide cross-couplings of the thermoelectric type.
\end{itemize}

In particular, we note that GK-type heat conduction discussed in
Section~\ref{sec:GK2} corresponds to a very simple Hamiltonian where we did not
use any coupling of the thermal vorticity $\F{}$ to other fields. In a more
general setting, more complicated regimes can be systematically explored by
considering more complex Hamiltonians that include the above types of couplings.

\subsubsection{Nonlinear connections and Finsler geometry}

As discussed in Section~\ref{ssec:gauge}, the two-temperature model has the
kinematics of an abelian gauge theory with linear connection. There are two
natural generalizations that allow to go beyond the setting of linear connection
and incorporate more complex physics. First, the proper temperature
\begin{equation*}
	\Tprop=\sqrt{-G^{\Asm\Bsm}\Jtot_\Asm \Jtot_\Bsm}
\end{equation*}
is a quadratic (Lorentzian) norm of the thermal impulse. In anisotropic or
strongly non-equilibrium media, it is natural to replace it by a general
function $\Tprop=\mathcal{F}(\Jtot_\Asm)$, positively homogeneous of degree one.
Such a function defines a Finsler (more precisely, Cartan) structure on the
cotangent bundle of $\mathscr{M}$. The associated geometry naturally carries a
\emph{nonlinear connection}, i.e. a connection that is not linear in the fiber
coordinates. Second, if the fiber of the bundle is not a vector space, e.g. if
the relevant thermal variable is $\ln\Ttot$, so that the structure group acts
multiplicatively on the temperature, the connection $\Jmicrovec$ is replaced by
a general Ehresmann connection. In both cases the field strength is no longer
given by the plain exterior derivative, and the resulting field equations become
nonlinear already at the kinematic level. For example, see
\cite{bejancu1988,saczuk1997,Clayton2021} for application of Finsler geometry in
solid mechanics and other fields. We leave these extensions for future work.

\subsection{Symmetric hyperbolicity of the extended model}\label{ssec:shtc.ext}

The two-temperature model, as well as the base model, belongs to the class of
symmetric hyperbolic thermodynamically compatible (SHTC) systems, e.g.
\cite{GodRom1996,Lucca2026,peshkov2026a} as long as the Hamiltonian is a convex
function of the state variables. The symmetric hyperbolic structure can be
easily seen in the dual variables, e.g. see \cite{GodRom1996,SHTC_GENERIC}.
Therefore, for simple Hamiltonians (e.g. the quadratic Hamiltonian
\eqref{eq:H.quadratic.GK2} with convex $\veps(\Stot)$ and positive parameters
$\Ch$, $\ch$, $c_\Smicro$, $\mu_0$, $\eps_0$), the two-temperature model has a
well-posed initial-value problem. However, once the model will be coupled with
other physical fields (e.g. mechanics, electromagnetism, etc.), the convexity of
the Hamiltonian has to be re-examined every time.

\subsection{Numerical discretization}\label{ssec:numerics}

Another importance of having a geometric structure for a model is that it has
direct consequences for its numerical discretization.

The constraints $\nabla\cdot\BFvec=0$ and $\BFvec=\alpha\nabla\times\Jmicrovec$
in \eqref{eq:inv.rescaled}, as well as $\nabla\times\Jtotvec=0$ in the
reversible case, are consequences of $\rmd^2=0$. In other words, they express
the exactness of the \emph{de Rham complex}. On the matter--time manifold
$\mathscr{M}$ this is the sequence of spaces of differential forms
$\Omega^0\xrightarrow{\rmd}\Omega^1\xrightarrow{\rmd}\Omega^2\xrightarrow{\rmd}\Omega^3\xrightarrow{\rmd}\Omega^4$
\cite{ArnoldFalkWinther2006}, to which $\varphi$, the impulses $\Jtotvec$ and
$\Jmicrovec$, and the thermal vorticity $\F{}$ belong. The involutions are
preserved exactly at the discrete level by schemes that reproduce this structure
by a \emph{discrete de Rham complex}. In such a complex, the discrete gradient,
curl and divergence map between discrete spaces of $0$-, $1$-, $2$- and
$3$-forms, satisfy $\mathrm{curl}_h\circ\mathrm{grad}_h=0$ and
$\mathrm{div}_h\circ\mathrm{curl}_h=0$ exactly, and commute with the
interpolation operators, e.g. see
\cite{Yee1966,EvansHawley1988,Nedelec1980,Bossavit1988,Hiptmair2002,
ArnoldFalkWinther2006,ArnoldFalkWinther2010,Desbrun2005,Lipnikov2014,dumbser2025a},
as well as compatible finite volume schemes for hyperbolic and SHTC systems
\cite{dumbser2025,Boscheri2024,dumbser2026}. In this way the differential
identities that follow from $\rmd^2=0$ hold exactly at the discrete level, and
divergence-cleaning techniques \cite{Dedneretal2002}, which are needed
otherwise, are avoided.

Since the model belongs to the SHTC class, structure-preserving schemes that
reproduce the energy conservation and the entropy inequality at the discrete
level (the so-called Hyperbolic Thermodynamically Compatible (HTC) schemes) are
available
\cite{HTC2022,Lucca2026,Boscheri2024,dumbser2025,ferrari2026,rinaldi2026}. They
exploit the same pairing of dynamical equations and kinematic identities that
underlies the geometric formulation. This is, in turn, the consequence of the
variational structure underlying the SHTC systems. Other variational
(multisymplectic) discretizations are available
\cite{Marsden2001,Gawlik2011,Morrison2017} and can be used to obtain schemes
with good long-time energy behavior in the ballistic regime, or as being a part
of a scheme for the whole dissipative system.

\subsubsection*{Declaration of generative AI and AI-assisted technologies}
During the preparation of this work, the author was using the AI assistant
Claude (Anthropic), with the model Claude Opus 5.5 to draft parts of the text,
to search and verify the literature, and to improve language, notation and
consistency. Deepreview services by Google was also used to check the
consistency of the manuscript. After using these tools, the author reviewed and
edited the text as needed. The author takes full responsibility for the content
of the published article.

\appendixpage
\begin{appendices}
\section{Coordinate-free form of the base model}\label{app:base.forms}

Here, we provide a formulation of the base model in the language of exterior
calculus. The thermal impulse is the $1$-form
\begin{equation}\label{eq:J.base.form}
	\Jtotvec = \rmd\varphi = \Jtot_\Asm\,\rmd X^\Asm 
	= -\Ttot\,\rmd T + \Jtot_\Ksm\,\rmd X^\Ksm .
\end{equation}
We denote by $\mathrm{vol}=\rmd T\wedge\rmd X^1\wedge\rmd X^2\wedge\rmd X^3$ the
volume $4$-form, and by $\mathrm{vol}_3=\rmd X^1\wedge\rmd X^2\wedge\rmd X^3$
its spatial part. We introduce the \emph{entropy four-current}, i.e. the vector
field
\begin{equation}\label{eq:entropy.current}
	\bm{\mathfrak{s}} := -\Lambda_{\Jtot_\Asm}\,\pd_\Asm
	= \Stot\,\pd_\Tsm + \Psi^\Ksm\,\pd_\Ksm ,
	\qquad
	\Psi^\Ksm = \Ham_{\Jtot_\Ksm},
\end{equation}
where we used $\Lambda_{\Jtot_\Osm}=-\Lambda_\Ttot=-\Stot$ and
$-\Lambda_{\Jtot_\Ksm}=\Ham_{\Jtot_\Ksm}$. By Cartan's formula, $\rmd(\iota_{\bm
v}\mathrm{vol})=(\pd_\Asm v^\Asm)\,\mathrm{vol}$ for any vector field $\bm v$,
and with $\iota$ being the interior product. The reversible base model
\eqref{eq:EL3} is therefore equivalent to
\begin{equation}\label{eq:base.forms}
	\rmd\left(\iota_{\bm{\mathfrak{s}}}\mathrm{vol}\right) = 0,
	\qquad
	\rmd\Jtotvec = 0 ,
\end{equation}
where
\begin{equation}
	\iota_{\bm{\mathfrak{s}}}\mathrm{vol}
	= \Stot\,\mathrm{vol}_3 - \rmd T\wedge\iota_{\JDtotvec}\mathrm{vol}_3
\end{equation}
is the entropy $3$-form: its spatial part is the entropy density, and its
mixed part is the entropy flux. Integrated over the boundary of any
four-dimensional region of $\mathscr{M}$, it gives the integral entropy balance
$\oint\iota_{\bm{\mathfrak{s}}}\mathrm{vol}=0$. In components, the second
equation in \eqref{eq:base.forms} is
\begin{equation}
	\rmd\Jtotvec
	= \left(\pd_\Tsm \Jtot_\Ksm + \pd_\Ksm\Ttot\right)\rmd T\wedge\rmd X^\Ksm
	+ \frac{1}{2}\left(\pd_\Ism \Jtot_\Ksm 
	- \pd_\Ksm \Jtot_\Ism\right)\rmd X^\Ism\wedge\rmd X^\Ksm = 0,
\end{equation}
i.e. the second equation in \eqref{eq:EL5} together with the spatial constraints
$\nabla\times\Jtotvec=0$. Similarly, the energy conservation \eqref{eq:cons.Ham}
reads $\rmd(\iota_{\bm{\mathfrak{e}}}\mathrm{vol})=0$ with the energy
four-current $\bm{\mathfrak{e}}:=\Ham\,\pd_\Tsm + q^\Ksm\,\pd_\Ksm$. Note that
\eqref{eq:base.forms} uses only the volume form and no metric. The metric enters
only through the Lagrangian, e.g. through the proper temperature $\Tprop =
\sqrt{-\Jtot_\Asm J^\Asm } = \sqrt{\Ttot^2 - \Jtot_\Ksm J^K} $.

With dissipation, the base model \eqref{eq:source} reads (recall that
$\F{\Ism\Ksm} := \pd_\Ism \Jtot_\Ksm - \pd_\Ksm \Jtot_\Ism$)
\begin{subequations}
	\begin{equation}\label{eq:base.forms.diss}
		\rmd\left(\iota_{\bm{\mathfrak{s}}}\mathrm{vol}\right) 
		= \mathscr{S}_\Osm\,\mathrm{vol},
		\qquad
		\F{} := \rmd\Jtotvec = \rmd T\wedge\bm{\mathcal{S}}
		+ \frac{1}{2}\F{\Ism\Ksm}\rmd X^\Ism\wedge\rmd X^\Ksm ,
	\end{equation}
	with
	\begin{equation}
		\bm{\mathcal{S}} 
		:= \mathscr{S}_\Ksm\,\rmd X^\Ksm 
		= -M_{\Ksm\Lsm}\Ham_{\Jtot_\Lsm}\,\rmd X^\Ksm.
	\end{equation}
\end{subequations}
Since $\F{}$ is exact, $\rmd\F{}=0$. In the $3+1$ split, this identity gives
\begin{equation}\label{eq:base.vorticity.production}
	\pd_\Tsm\F{\Ism\Ksm} = \pd_\Ism \mathscr{S}_\Ksm - \pd_\Ksm \mathscr{S}_\Ism ,
	\qquad
	\text{or }
	\qquad
	\pd_\Tsm(\nabla\times\Jtotvec) = \nabla\times\bm{\mathcal{S}},
\end{equation}
i.e. the dissipation produces the spatial thermal vorticity.

\end{appendices}

\printbibliography

\end{document}